\documentclass[aps,pra,twocolumn,superscriptaddress,showpacs,nofootinbibfloatfix,amsmath,amsfonts,amssymb]{revtex4-2}%
\usepackage{amsmath,amsfonts,amssymb,color}
\usepackage{amsthm}
\usepackage{leftidx}
\usepackage{graphicx}
\usepackage{xcolor}
\usepackage{dcolumn}
\usepackage{bm}
\usepackage{epstopdf}
\usepackage{epsfig}
\usepackage{environ}
\usepackage{pdfcomment}

\usepackage{multirow}
\usepackage{setspace}
\usepackage{color}

\usepackage{float}
\usepackage[T1]{fontenc}
\usepackage[latin9]{inputenc}
\usepackage{setspace}
\usepackage{esint}

\usepackage{wasysym}

\begin{document}


\title{Topological characterization of phase transitions and critical edge
	states in one-dimensional non-Hermitian systems with sublattice symmetry}

\author{Longwen Zhou}
\email{zhoulw13@u.nus.edu}
\affiliation{%
	College of Physics and Optoelectronic Engineering, Ocean University of China, Qingdao, China 266100
}
\affiliation{%
	Key Laboratory of Optics and Optoelectronics, Qingdao, China 266100
}
\affiliation{%
	Engineering Research Center of Advanced Marine Physical Instruments and Equipment of MOE, Qingdao, China 266100
}
\author{Rujia Jing}
\affiliation{%
	College of Physics and Optoelectronic Engineering, Ocean University of China, Qingdao, China 266100
}
\author{Shenlin Wu}
\affiliation{%
	College of Physics and Optoelectronic Engineering, Ocean University of China, Qingdao, China 266100
}

\date{\today}

\begin{abstract}
Critical edge states appear at the bulk gap closing points of topological
transitions. Their emergence signify the existence of topologically
nontrivial critical points, whose descriptions fall outside the scope
of gapped topological matter. In this work, we reveal and characterize
topological critical points and critical edge states in non-Hermitian
systems. By applying the Cauchy's argument principle to two characteristic
functions of a non-Hermitian Hamiltonian, we obtain a pair of winding
numbers, whose combination yields a complete description of gapped
and gapless topological phases in one-dimensional, two-band non-Hermitian
systems with sublattice symmetry. Focusing on a broad class of non-Hermitian
Su-Schrieffer-Heeger chains, we demonstrate the applicability of our
theory for characterizing gapless symmetry-protected topological phases,
topologically distinct critical points, phase transitions along non-Hermitian
phase boundaries and their associated topological edge modes. Our
findings not only generalize the concepts of topologically nontrivial
critical points and critical edge modes to non-Hermitian setups, but
also yield additional insights for analyzing topological transitions
and bulk-edge correspondence in open systems.
\end{abstract}

\maketitle

\section{Introduction\label{sec:Int}}

Non-Hermitian physics has attracted great attention in the past decades
(see \cite{NHRev01,NHRev02,NHRev03,NHRev04,NHRev05,NHRev06,NHRev07,NHRev08,NHRev09,NHRev10,NHRev11,NHRev12,NHRev13,NHRev14,NHRev15,NHRev16,NHRev17}
for reviews). Unique non-Hermitian phenomena, such as the exceptional
point \cite{EP01,EP02,EP03,EP04,EP05,EP06}, non-Hermitian skin effect
\cite{NHSE01,NHSE02,NHSE03,NHSE04,NHSE05,NHSE06} and enriched classification
of topological matter \cite{NHTC01,NHTC02,NHTC03,NHTC04,NHTC05,NHTC06}
have been discovered theoretically and explored further in experiments,
yielding new perspectives on sensing, wave-guiding and other device
applications \cite{NHAP01,NHAP02,NHAP03,NHAP04,NHAP05,NHAP06,NHAP07}.
In the study of non-Hermitian topological matter, the focus is mainly
rested on systems whose nontrivial topology is associated to a point
or a line spectral gap, in which symmetry-protected topological edge
states reside \cite{NHRev05}. When the bulk spectral gap closes,
a phase transition is expected to occur. There, standard topological
markers of gapped phases like the winding and Chern numbers become
ill-defined. The fates of edge modes related to these topological
invariants in nearby gapped phases would also be unclear at the exact
transition point. The possibility of finding edge states
rooted in the nontrivial topology of non-Hermitian gapless
critical points has yet to be unveiled.

Recently, the study of symmetry-protect topological (SPT) phases has
been extended to gapless  situations~\cite{gSPT01,gSPT02,gSPT03,gSPT04,gSPT05,gSPT06,gSPT07,gSPT08,gSPT09,gSPT10,gSPT11,gSPT12,gSPT13,gSPT14,gSPT142,gSPT15,gSPT16,gSPT17,gSPT18,gSPT19,gSPT20,gSPT201,gSPT202,gSPT21,gSPT22,gSPT23,gSPT24,gSPT25,gSPT26,gSPT27,gSPT28,gSPT29,gSPT30,gSPT31,gSPT32,gSPT322,gSPT34,gSPT35,gSPT36,gSPT37,gSPT38,gSPT39,gSPT40,gSPT41,gSPT43}.
The theoretical description of a gapless SPT (gSPT) state not only
goes beyond the Landau paradigm of phase transitions based on local
order parameters and symmetry-breaking \cite{LandauBook}, but also
falls outside the standard classification of topological insulators
and superconductors following the tenfold way \cite{Tenfold}. In
the simplest case, a gSPT phase can be realized by a critical point
between two topologically nontrivial phases \cite{gSPT26}. In one
spatial dimension, such a realization could be the transition point
between two gapped phases with different and nonzero topological invariants
in a spin or superconducting chain \cite{gSPT04}, and the nontrivial
topology of the critical point is signified by the presence of degenerate
edge modes at the gap-closing point of the bulk energy spectrum \cite{gSPT11}.
Be coexisting with a gapless bulk, these topological edge modes can
also be regarded as critical. In the presence of non-Hermitian effects,
the description of gSPT phases could become more complicated. On the
one hand, unique non-Hermitian topology related to the exceptional
point and non-Bloch band theory may lead to new forms of gSPT phases
beyond closed-system constraints. On the other hand, gain and loss
or nonreciprocal effects may influence the existence and stability
of critical edge modes in an originally Hermitian gSPT phase. Addressing
these issues could not only extend the scope of gSPT phases to
non-Hermitian setups, but also offer potential insights for further
explorations of mixed-state topology \cite{MSTP01,MSTP02,MSTP03,MSTP04,MSTP05,MSTP06}
and bulk-edge correspondence in gapless open systems.

\begin{figure*}
	\begin{centering}
		\includegraphics[scale=0.29]{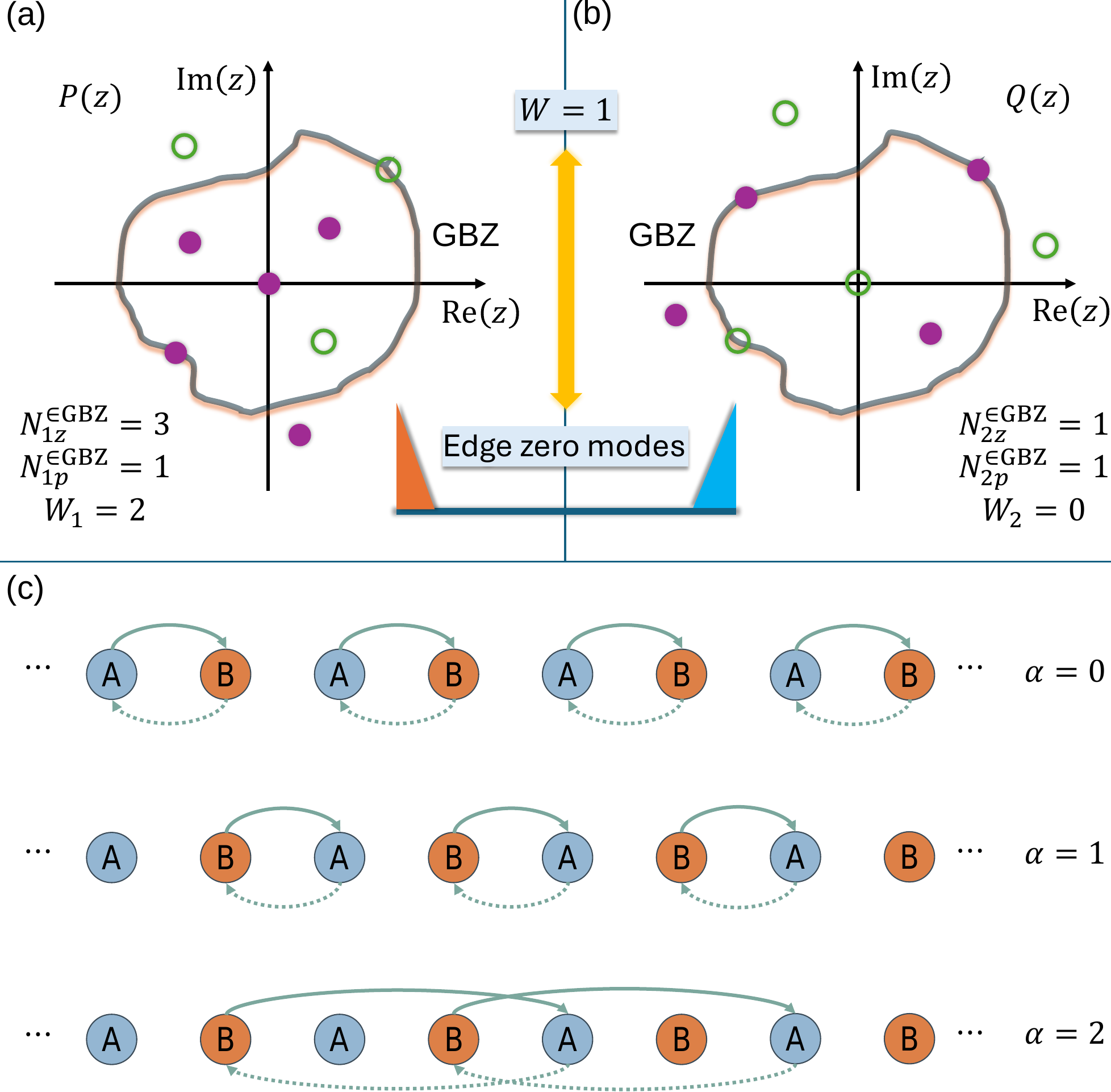}
		\par\end{centering}
	\caption{Schematic diagrams of the theory and model. (a) and (b) illustrate the configurations of zeros (in purple dots) and poles (in green circles) of the characteristic functions $P(z)$ and $Q(z)$ [Eq.~(\ref{eq:PzQz})] on the complex $z$ plane, with the GBZ given by the sinuous contours. Only zeros/poles inside the GBZ contribute to the winding numbers $W_1$ and $W_2$ [Eq.~(\ref{eq:W12})], whose combination predicts the topological invariant $W=1$ [Eq.~(\ref{eq:W})] and number of edge zero modes $N_0=2$ [Eq.~(\ref{eq:N0})] for the case illustrated in (a)--(b).
	(c) shows NHSSH $\alpha$ chains with $\alpha=0,1,2$.
	In each case, the solid and dotted curves with arrows denote hopping
	amplitudes from left to right and right to left between sublattices
	A (cyan balls) and B (orange balls). \label{fig:Sketch}}
\end{figure*}

In this work, we propose a theoretical framework to characterize topologically
nontrivial critical points and critical edge states in one-dimensional
(1D), two-band non-Hermitian systems. Our approach is applicable to
the description of both gapped and gapless sublattice-symmetry-protected
topological phases. Moreover, it does not concern whether the underlying
non-Hermitian band theory is in Bloch (with standard Brillouin zone)
or non-Bloch (with generalized Brillouin zone) form. The rest of the
paper is organized as follow. In Sec.~\ref{sec:The}, we formulate
the definition of topological invariants and the related bulk-edge
correspondence in our theory. The topological invariants are deduced
from the algebraic zero-pole counting of two characteristic polynomials
for a given non-Hermitian Hamiltonian [Figs.~\ref{fig:Sketch}(a)--\ref{fig:Sketch}(b)]. In Sec.~\ref{sec:Res}, we
apply our theory to characterize the topological phases, phase transitions
and edge states in a broad class of non-Hermitian Su-Schrieffer-Heeger
(NHSSH) chains, with a focus on the topological nature and bulk-edge
correspondence at their gapless critical points. Different classes
of topologically distinct critical points and phase transitions along
topological phase boundaries are revealed, and the configuration of
their critical edge modes are identified. In Sec.~\ref{sec:Sum},
we summarize our results and discuss potential future studies.
Some further theoretical details and model illustrations
are presented in Appendices \ref{App:Theory}--\ref{sec:App1}.

\section{Theory\label{sec:The}}

In this section, we outline the theoretical framework we proposed
to characterize gapped and gapless topological phases in 1D, two-band
non-Hermitian systems with sublattice symmetry. Despite showing the
basic formulation, we also discuss the issue of some previous
theories on the topological characterization of phase transition points
with vanishing bulk gaps and critical edge states. The application
of our theory to concrete models is presented in the next section. 
Some further theoretical details are provided in Appendix \ref{App:Theory}.

Under an appropriate basis choice, the generic Hamiltonian of a 1D,
two-band, sublattice-symmetric non-Hermitian model can be expressed
in an off-diagonal form as 
\begin{equation}
	H(k)=\begin{bmatrix}0 & f(k)\\
		g(k) & 0
	\end{bmatrix},\label{eq:Hk}
\end{equation}
where $k\in[-\pi,\pi]$ is the quasimomentum.
The $H(k)$ in Eq.~(\ref{eq:Hk}) possesses the sublattice symmetry $\Gamma=\sigma_z$. Its phases at half-filling are then characterized by integer-valued topological numbers. To consistently define these invariants for both gapped and gapless phases, we introduce a pair of characteristic functions for the complex-extended Hamiltonian $H(z)\equiv H(k\rightarrow-i\ln{z})$, which are given by [see Appendix \ref{App:Theory} for further details and Figs.~\ref{fig:Sketch}(a)--\ref{fig:Sketch}(b) for graphic illustrations]
\begin{equation}
	P(z)=\frac{g(z)}{f(z)},\qquad Q(z)=f(z)g(z),\label{eq:PzQz}
\end{equation}
where $f(z)\equiv f(k\rightarrow-i\ln z)$ and $g(z)\equiv g(k\rightarrow-i\ln z)$ are ``complex continuations'' of the $f(k)$ and $g(k)$ in $H(k)$ to the whole complex plane $z\in\mathbb{C}$. Inside a closed contour on the complex plane, the difference
between the numbers of zeros and poles (including their multiplicities
and orders) of $P(z)$ and $Q(z)$ yield two integer winding
numbers $W_{1}$ and $W_{2}$, respectively, according to the Cauchy's
argument principle. Our theoretical proposal is based on the topological
messages encoded in these winding numbers.

In previous studies \cite{NHSE04,NHSE07}, it was found that by identifying
the first $2\mu=2\max\{m,n\}$ zeros and/or poles of $P(z)$ (ordered
according to their magnitudes), we would obtain a winding number $W_{0}\equiv N_{z}^{\in2\mu}-N_{p}^{\in2\mu}$,
which could characterize each topological phase of a 1D, two-band,
sublattice-symmetric non-Hermitian system and determine the number
$N_{0}$ of its edge zero modes via the relation $N_{0}=|W_{0}|$.
Here, $m$ and $n$ are given by the highest negative powers of $f(z)$
and $g(z)$ in $z$. $N_{z}^{\in2\mu}$ and $N_{p}^{\in2\mu}$ are
the numbers of zeros and poles of $P(z)$ inside the set of its first
$2\mu$ zeros/poles. This algebraic approach is proved to be equivalent
to other methods based on the generalized Brillouin zone (GBZ) and
non-Bloch band theory for gapped phases \cite{NHSE02,NHSE03,NHSE07}.
However, this $P(z)$-based theory may lead to ambiguous predictions
about the non-Hermitian topology and edge states at phase transition
points, where the Bloch or non-Bloch bands of the system are touched
at zero energy. We discuss an explicit example to showcase this ambiguity in Appendix \ref{App:Theory}.

At critical points, a key issue of the existing theory is that there could be competing zeros
and poles of equal magnitudes in $f(z)$ and $g(z)$, whose topological signatures
may subject to cancellations in $P(z)$ by definition. Their contributions
are instead all retained in the function $Q(z)$
at the gap closing point. Therefore, we propose to introduce
an additional winding number by applying the Cauchy's argument principle
to $Q(z)$. More precisely, by taking GBZ as the integration contour, the winding numbers of $P(z)$ and $Q(z)$ are given by
\begin{alignat}{1}
	W_{1}& \equiv N_{1z}^{\in{\rm GBZ}}-N_{1p}^{\in{\rm GBZ}},\nonumber \\
	W_{2}& \equiv N_{2z}^{\in{\rm GBZ}}-N_{2p}^{\in{\rm GBZ}}.\label{eq:W12}
\end{alignat}
Here, $N_{1z}^{\in{\rm GBZ}}$ and $N_{1p}^{\in{\rm GBZ}}$ ($N_{2z}^{\in{\rm GBZ}}$
and $N_{2p}^{\in{\rm GBZ}}$) denote the numbers of zeros and poles
of $P(z)$ ($Q(z)$) inside the GBZ. Note that the definition of $W_{1}$ here
is consistent with that of $W_{0}$ in the existing theory
\cite{NHSE07}. Combining the information of $W_{1}$ and
$W_{2}$, we arrive at the final topological invariant $W$ and bulk-edge
correspondence of 1D, two-band systems with sublattice symmetry, i.e.,
\begin{alignat}{1}
	W& = \frac{W_{1}+W_{2}}{2},\label{eq:W}\\
	N_{0}& = |W_{1}+W_{2}|=2|W|,\label{eq:N0}
\end{alignat}
where $N_{0}$ denotes the total number of edge zero modes.

In Appendix \ref{App:Theory}, we illustrate the applicability of our theory to the simplest SSH model away from and at its criticality. In the next
section, we utilize our theory to depict the topology and edge
states in both gapped phases and along gapless phase boundaries for
a broad class of sublattice-symmetric non-Hermitian models.

\section{Results\label{sec:Res}}

In this section, we demonstrate the applicability of our theory to
the characterization of topologically nontrivial critical points and
zero-energy edge modes in non-Hermitian systems. We first introduce
a general class of 1D non-Hermitian models with sublattice symmetry
in Sec.~\ref{subsec:Mod}, which will be referred to as NHSSH $\alpha$
chains. Our theory is then applied to describing gapped topological
phases, distinguishing between topologically trivial and nontrivial
critical points, and depicting topological phase transitions without
gap closing/reopening in Secs.~\ref{subsec:SSHaa}
and \ref{subsec:SSHaaa}. Theoretical analyses are complemented by
numerical calculations of the spectra and edge states in each case.
For completeness, we also discuss in Appendix \ref{sec:SSHa} the topological phases for the simplest class of NHSSH $\alpha$ chains introduced in Sec.~\ref{subsec:Mod}, which could be either fully gapped or gapless without having topological phase transitions.

\subsection{Model\label{subsec:Mod}}
The generic NHSSH $\alpha$ chain is formally described by the Hamiltonian
$\hat{H}=\sum_{\alpha\in\mathbb{Z}}\hat{H}_{\alpha}$, where 
\begin{equation}
	\hat{H}_{\alpha}=\sum_{n\in\mathbb{Z}}\left(J_{\alpha}^{{\rm L}}\hat{b}_{n}^{\dagger}\hat{a}_{n+\alpha}+J_{\alpha}^{{\rm R}}\hat{a}_{n+\alpha}^{\dagger}\hat{b}_{n}\right),\quad J_{\alpha}^{{\rm L}},J_{\alpha}^{{\rm R}}\in\mathbb{C}.\label{eq:Ha}
\end{equation}
$\hat{a}_{n}^{\dagger}$ ($\hat{b}_{n}^{\dagger}$) creates a
particle in the sublattice A (B) of unit cell $n$. $J_{\alpha}^{{\rm L}}$
($J_{\alpha}^{{\rm R}}$) denotes the directional hopping amplitude from sublattices
A to B (B to A) over $\alpha$ lattice sites. The $\hat{H}$
is non-Hermitian whenever there exists an $\alpha$ such that $J_{\alpha}^{{\rm L}}\neq(J_{\alpha}^{{\rm R}})^{*}$.
The Hermitian SSH model is recovered by setting $\hat{H}=\sum_{\alpha=0,1}\hat{H}_{\alpha}$
and $J_{\alpha}^{{\rm L}}=J_{\alpha}^{{\rm R}}=J_{\alpha}\in\mathbb{R}$.
The lattice geometries of $\hat{H}_{\alpha}$ with
$\alpha=0,1,2$ are shown separately in Fig.~\ref{fig:Sketch}(c).

Under the periodic boundary condition (PBC), we can express the Hamiltonian
$\hat{H}$ in momentum space via Fourier transformations
\begin{equation}
	\hat{a}_{n}=\frac{1}{\sqrt{N}}\sum_{k}e^{ikn}\hat{a}_{k},\quad\hat{b}_{n}=\frac{1}{\sqrt{N}}\sum_{k}e^{ikn}\hat{b}_{k},\label{eq:FFT}
\end{equation}
where $N$ is the total number of unit cells and $k\in[-\pi,\pi]$. In momentum space, the component
$\hat{H}_{\alpha}$ has the form $\hat{H}_{\alpha}=\sum_{k}\hat{\Psi}_{k}^{\dagger}H_{\alpha}(k)\hat{\Psi}_{k}$,
where $\hat{\Psi}_{k}^{\dagger}\equiv(\hat{a}_{k}^{\dagger},\hat{b}_{k}^{\dagger})$.
The $H_{\alpha}(k)$ is a $2\times2$ off-diagonal matrix, given by
\begin{equation}
	H_{\alpha}(k)=\begin{bmatrix}0 & J_{\alpha}^{{\rm R}}e^{-ik\alpha}\\
		J_{\alpha}^{{\rm L}}e^{ik\alpha} & 0
	\end{bmatrix}.\label{eq:Hak}
\end{equation}
It is clear that each $H_{\alpha}(k)$ {[}and also the combined Bloch
Hamiltonian $H(k)=\sum_{\alpha\in\mathbb{Z}}H_{\alpha}(k)${]} has
the sublattice symmetry $\Gamma=\sigma_{z}$, in the sense that $\Gamma H_{\alpha}(k)\Gamma=-H_{\alpha}(k)$
and $\Gamma^{2}=\sigma_0$ for any $\alpha$. The theoretical framework developed
in the last section is thus applicable. Making the substitution $e^{ik}\rightarrow z$
with $z\in\mathbb{C}$, we find the extension of $H_{\alpha}(k)$
to the whole complex plane as
\begin{equation}
	H_{\alpha}(z)=\begin{bmatrix}0 & J_{\alpha}^{{\rm R}}z^{-\alpha}\\
		J_{\alpha}^{{\rm L}}z^{\alpha} & 0
	\end{bmatrix}.\label{eq:Haz}
\end{equation}
The extended Bloch Hamiltonian of the NHSSH $\alpha$ chain is then
given by $H(z)=\sum_{\alpha\in\mathbb{Z}}H_{\alpha}(z)$.
The topological properties of $H_{\alpha}(z)$ with a fixed $\alpha$ is discussed in Appendix \ref{sec:SSHa}.

In the following, we consider two different types of NHSSH $\alpha$ chains, and apply our theory to characterize their topological phases and bulk-edge correspondence.

\subsection{$(\alpha,\alpha')$ chain: topologically trivial/nontrivial critical
	points\label{subsec:SSHaa}}

The NHSSH $(\alpha,\alpha')$ chain has the complex-extended Hamiltonian
$H(z)=H_{\alpha}(z)+H_{\alpha'}(z)$, where $\alpha\neq\alpha'$.
It has two competing length scales in its hopping amplitudes $J_{\alpha}^{{\rm L,R}}$
and $J_{\alpha'}^{{\rm L,R}}$, making it possible to find topological
phase transitions. We assume $0\leq\alpha<\alpha'$ throughout this
subsection. The cases with $\alpha,\alpha'\leq0$ can be treated following
a similar routine.

The theory outlined in Sec.~\ref{sec:The} allows us to identify
$f(z)=J_{\alpha}^{{\rm R}}z^{-\alpha}+J_{\alpha'}^{{\rm R}}z^{-\alpha'}$
and $g(z)=J_{\alpha}^{{\rm L}}z^{\alpha}+J_{\alpha'}^{{\rm L}}z^{\alpha'}$
for the $(\alpha,\alpha')$ chain. The characteristic functions $P(z)$
and $Q(z)$ are thus given by
\begin{alignat}{1}
	P(z)& = ({J_{\alpha}^{{\rm L}}z^{\alpha}+J_{\alpha'}^{{\rm L}}z^{\alpha'}})/({J_{\alpha}^{{\rm R}}z^{-\alpha}+J_{\alpha'}^{{\rm R}}z^{-\alpha'}}),\label{eq:Paaz}\\
	Q(z)& = (J_{\alpha}^{{\rm L}}z^{\alpha}+J_{\alpha'}^{{\rm L}}z^{\alpha'})(J_{\alpha}^{{\rm R}}z^{-\alpha}+J_{\alpha'}^{{\rm R}}z^{-\alpha'}).\label{eq:Qaaz}
\end{alignat}
It is straightforward to inspect that $P(z)$ has a zero of order
$(\alpha'+\alpha)$ at $z=0$, other $(\alpha'-\alpha)$ zeros of
order $1$ along the circle of radius $r^{{\rm L}}=|J_{\alpha}^{{\rm L}}/J_{\alpha'}^{{\rm L}}|^{1/(\alpha'-\alpha)}$,
and $(\alpha'-\alpha)$ poles of order $1$ along the circle of radius
$r^{{\rm R}}=|J_{\alpha'}^{{\rm R}}/J_{\alpha}^{{\rm R}}|^{1/(\alpha'-\alpha)}$.
On the other hand, $Q(z)$ has $(\alpha'-\alpha)$ zeros of order
$1$ along the circle of radius $r^{{\rm L}}$, other $(\alpha'-\alpha)$
zeros of order $1$ along the circle of radius $r^{{\rm R}}$, and
a pole of order $(\alpha'-\alpha)$ at $z=0$. The locations of these
zeros and poles relative to the GBZ fully determine the topology of
gapped phases and gapless phase boundaries of the $(\alpha,\alpha')$
chain. 

The GBZ of NHSSH $(\alpha,\alpha')$ chain can be obtained from the
two middle solutions of the equation $Q(z)=Q(ze^{i\theta})$ for $\theta\in[0,2\pi]$
\cite{NHSE03}. Direct calculations yield the solution $z^{2(\alpha'-\alpha)}=J_{\alpha}^{{\rm L}}J_{\alpha'}^{{\rm R}}e^{-i(\alpha'-\alpha)\theta}/(J_{\alpha'}^{{\rm L}}J_{\alpha}^{{\rm R}})$.
Therefore, the GBZ of $(\alpha,\alpha')$ chain has a circular shape with radius
\begin{equation}
	r_{0}=\left|{J_{\alpha}^{{\rm L}}J_{\alpha'}^{{\rm R}}}/({J_{\alpha'}^{{\rm L}}J_{\alpha}^{{\rm R}}})\right|^{\frac{1}{2(\alpha'-\alpha)}}.\label{eq:GBZaa}
\end{equation}
Under the non-Hermitian condition $J_{\alpha}^{{\rm L}}\neq(J_{\alpha}^{{\rm R}})^{*}$
and/or $J_{\alpha'}^{{\rm L}}\neq(J_{\alpha'}^{{\rm R}})^{*}$, we
generally have $r_{0}\neq1$, making the GBZ different from the
standard BZ of a Hermitian model that corresponds to a unit circle
on the complex plane. The critical (gap-closing) points of the NHSSH
$(\alpha,\alpha')$ chain can be obtained by inserting the GBZ solution
$z_{0}=r_{0}e^{i\phi}$ ($\phi\in[0,2\pi]$) into the equality $Q(z)=[E(z)]^{2}=0$,
yielding the phase boundary in the hopping parameter space
\begin{equation}
	|{\cal J}_{\alpha'}|=|{\cal J}_\alpha|,\qquad{\cal J}_\alpha\equiv{J_{\alpha}^{{\rm L}}J_{\alpha}^{{\rm R}}}\quad\forall\alpha.\label{eq:PBaa}
\end{equation}
The parameter regions $|{\cal J}_{\alpha'}|<|{\cal J}_\alpha|$
and $|{\cal J}_{\alpha'}|>|{\cal J}_\alpha|$
are then expected to belong to distinct topological phases.

We next determine the locations of zeros/poles of $P(z)$ and $Q(z)$
relative to the GBZ. Using the relevant radii $r^{{\rm L}}$
and $r^{{\rm R}}$,
we obtain
\begin{equation}
	\frac{r^{{\rm L}}}{r_{0}}=\left|{{\cal J}_\alpha}/{{\cal J}_{\alpha'}}\right|^{\frac{1}{2(\alpha'-\alpha)}},\quad\frac{r^{{\rm R}}}{r_{0}}=\left|{{\cal J}_{\alpha'}}/{{\cal J}_{\alpha}}\right|^{\frac{1}{2(\alpha'-\alpha)}}.\label{eq:Raa}
\end{equation}
Therefore, if $|{{\cal J}_{\alpha'}}|<|{{\cal J}_{\alpha}}|$,
the zeros/poles lying along the circles of radii $r^{{\rm L}}$ and
$r^{{\rm R}}$ are outside and inside the GBZ, respectively. On the
contrary, when $|{{\cal J}_{\alpha'}}|>|{{\cal J}_{\alpha}}|$,
the zeros/poles lying along the circles of radii $r^{{\rm L}}$ and
$r^{{\rm R}}$ are inside and outside the GBZ, respectively. If $|{{\cal J}_{\alpha'}}|=|{{\cal J}_{\alpha}}|$,
we have $r^{{\rm L}}=r^{{\rm R}}=r_{0}$, which means that the zeros/poles
along the circles of radii $r^{{\rm L}}$ and $r^{{\rm R}}$ are
both on the GBZ when the phase boundary is reached. Finally, the zeros
and poles at $z=0$ are all inside the GBZ so long as the GBZ radius
$r_{0}>0$. 

We could now obtain the winding numbers $W_{1}$ and $W_{2}$ of $P(z)$
and $Q(z)$ following the Cauchy's argument principle. Specially,
we find from the Eq.~(\ref{eq:W12}) that 
\begin{equation}
	(W_{1},W_{2})=\begin{cases}
		(2\alpha,0), & |{{\cal J}_{\alpha'}}|<|{{\cal J}_{\alpha}}|\\
		(\alpha+\alpha',\alpha-\alpha'), & |{{\cal J}_{\alpha'}}|=|{{\cal J}_{\alpha}}|\\
		(2\alpha',0), & |{{\cal J}_{\alpha'}}|>|{{\cal J}_{\alpha}}|
	\end{cases},\label{eq:W12aa}
\end{equation}
Put together, we arrive at the following topological invariant and
bulk-edge correspondence of the NHSSH $(\alpha,\alpha')$ chain assuming
$0\leq\alpha<\alpha'$ for any $\alpha,\alpha'$, i.e.,
\begin{alignat}{1}
	W&=\frac{W_{1}+W_{2}}{2}=\begin{cases}
		\alpha, & |{\cal J}_{\alpha'}|\leq|{\cal J}_{\alpha}|\\
		\alpha', & |{\cal J}_{\alpha'}|>|{\cal J}_{\alpha}|
	\end{cases},\label{eq:Waa}\\
	N_{0}&=2|W|=\begin{cases}
	2\alpha, & |{\cal J}_{\alpha'}|\leq|{\cal J}_{\alpha}|\\
	2\alpha', & |{\cal J}_{\alpha'}|>|{\cal J}_{\alpha}|
	\end{cases}.\label{eq:BBCaa}
\end{alignat}
We conclude that the critical points along $|{\cal J}_{\alpha'}|=|{\cal J}_{\alpha}|$
are topologically nontrivial whenever $\alpha>0$. Their nontrivial
topologies are characterized by a quantized winding number $W=\alpha$
and a number $N_{0}=2\alpha$ of edge modes coexisting
with a gapless bulk at zero energy. Instead, the critical points along
$|{\cal J}_{\alpha'}|=|{\cal J}_{\alpha}|$
are topologically trivial when $\alpha=0$, regardless of the value
of $\alpha'$. Away from the critical points, we find two distinct
gapped phases with the winding numbers $\alpha,\alpha'$ and the numbers
of edge zero modes $2\alpha,2\alpha'$. These gapped phases and gapless
phase boundaries could thus
be completely described within our theoretical framework.

We now verify our theory with two sets of computational examples.
One of them shows trivial critical points, while the other holds
topologically nontrivial critical points with edge zero modes. We
first consider the case with $(\alpha,\alpha')=(0,1)$. The complex-extended
Hamiltonian reads $H_{01}(z)\equiv H_{0}(z)+H_{1}(z)$. For the hopping
parameters, we choose $J_{0}^{{\rm L}}=J-\gamma$, $J_{0}^{{\rm R}}=J+\gamma$
($J,\gamma\in\mathbb{R}$), and set $J_{1}^{{\rm L}}=J_{1}^{{\rm R}}=1$
as the unit of energy. This $H_{01}(z)$ is non-Hermitian
when $\gamma\neq0$. The non-Hermiticity is originated
from asymmetric intracell hoppings. Following Eqs.~(\ref{eq:GBZaa})
and (\ref{eq:PBaa}), the radius $r_{0}$ of GBZ and the phase boundary
equation in this case are given by
\begin{equation}
	r_{0}=\sqrt{|J-\gamma|/|J+\gamma|},\qquad J^{2}=\gamma^{2}\pm1.\label{eq:r0PBaa}
\end{equation}
The topological winding number $W$ and the number of zero-energy
edge modes $N_{0}$ are further predicted by Eqs.~(\ref{eq:Waa})
and (\ref{eq:BBCaa}) as
\begin{alignat}{1}
	W&=\begin{cases}
		1, & |J^{2}-\gamma^{2}|<1\\
		0, & {\rm Otherwise}
	\end{cases},\label{eq:Waa1}\\
	N_{0}&=\begin{cases}
		2, & |J^{2}-\gamma^{2}|<1\\
		0, & {\rm Otherwise}
	\end{cases}.\label{eq:BBCaa1}
\end{alignat}
We notice that in the limit $\gamma=0$, the properties
of Hermitian SSH model are recovered \cite{TPBook}.

The Eqs.~(\ref{eq:r0PBaa})--(\ref{eq:BBCaa1}) can be verified by computing the zero energy
solutions of the system explicitly. In lattice representation,
the Hamiltonian of NHSSH $(0,1)$ chain reads $\hat{H}_{01}=\hat{H}_{0}+\hat{H}_{1}$,
where $\hat{H}_{0}=\sum_{n}[(J+\gamma)\hat{a}_{n}^{\dagger}\hat{b}_{n}+(J-\gamma)\hat{b}_{n}^{\dagger}\hat{a}_{n}]$
and $\hat{H}_{1}=\sum_{n}(\hat{b}_{n}^{\dagger}\hat{a}_{n+1}+\hat{a}_{n+1}^{\dagger}\hat{b}_{n})$.
For a chain with unit cell indices $n=1,2,...,N$ under the open boundary condition (OBC), the (non-normalized) zero-energy eigenstates can be
found in the limit $N\rightarrow\infty$ in symmetrized basis as (see Appendix \ref{sec:App1}
for details)
\begin{alignat}{1}
	|\psi_{0}^{{\rm L}}\rangle= & \sum_{n=1}^{N}[-\varrho(J-\gamma)]^{n-1}\hat{a}_{n}^{\dagger}|\emptyset\rangle,\nonumber \\
	|\psi_{0}^{{\rm R}}\rangle= & \sum_{n=1}^{N}[-\varrho^{-1}(J+\gamma)]^{N-n}\hat{b}_{n}^{\dagger}|\emptyset\rangle,\label{eq:H01Edge}
\end{alignat}
where $|\varrho|=\sqrt{|J+\gamma|/|J-\gamma|}$. These states could
describe a pair of edge zero modes if and only if $|J^{2}-\gamma^{2}|<1$
(see also Appendix \ref{sec:App1}), which is exactly the condition
for the system to be in its gapped topological phase with the winding
number $W=1$ and the number of zero-energy edge modes $N_{0}=2$,
as reported in Eqs.~(\ref{eq:Waa1}) and (\ref{eq:BBCaa1}). Our theoretical
predictions are thus confirmed. Notably, the edge zero modes $|\psi_{0}^{{\rm L}}\rangle$
and $|\psi_{0}^{{\rm R}}\rangle$ become delocalized at the critical
points $J^{2}-\gamma^{2}=\pm1$. The gapless phase boundary of the
NHSSH $(0,1)$ chain is thus topologically trivial, which is characterized
by the winding number $W=0$ and the number of edge zero modes $N_{0}=0$.

\begin{figure}
	\begin{centering}
		\includegraphics[scale=0.5]{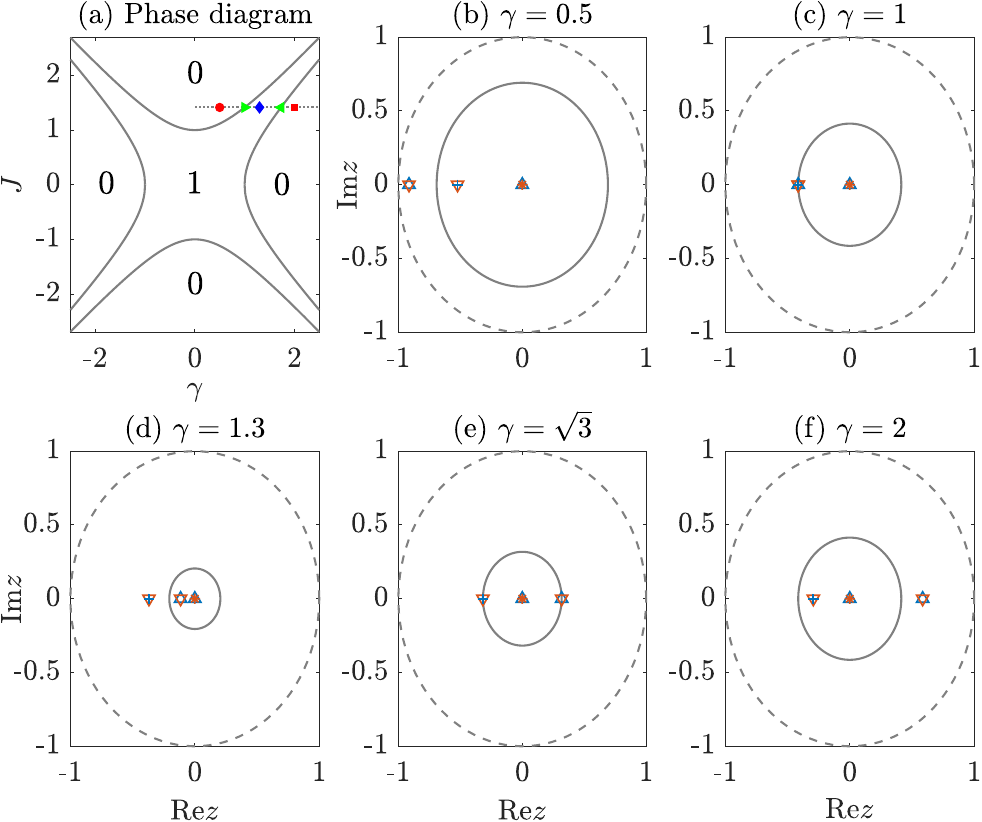}
		\par\end{centering}
	\caption{Phase diagram, GBZ and zero/pole distributions of the NHSSH $(0,1)$
		chain. (a) Topological phase diagram, with the values of
		$W$ shown explicitly in each phase. Phase boundaries are given by
		the gray solid lines. Along the dotted line, the symbols $\CIRCLE$,
		$\blacktriangleright$, $\blacklozenge$, $\blacktriangleleft$, and
		$\blacksquare$ highlight the points in the parameter space with $J=\sqrt{2}$
		and $\gamma=0.5,1,1.3,\sqrt{3}$, and $2$, respectively. (b)--(f)
		show the BZ (in dashed rings), GBZ (in solid rings), zeros
		of $P(z)$ (in $\vartriangle$), poles of $P(z)$ (in $+$), zeros
		of $Q(z)$ (in $\triangledown$), and poles of $Q(z)$ (in $*$) for
		the five points in (a) from left to right on the complex
		$z$-plane, respectively. \label{fig:PD1}}
\end{figure}

Below, we provide further evidence to testify our theoretical discoveries.
Plugging $(J_{0}^{{\rm L}},J_{0}^{{\rm R}})=(J-\gamma,J+\gamma)$
and $(J_{1}^{{\rm L}},J_{1}^{{\rm R}})=(1,1)$ into the Eqs.~(\ref{eq:Paaz})
and (\ref{eq:Qaaz}), we find that the $P(z)$ has two zeros at $z=0,-(J-\gamma)$
and a pole at $z=-1/(J+\gamma)$, while the $Q(z)$ has two zeros
at $z=-(J-\gamma),-1/(J+\gamma)$ and a pole at $z=0$. All these
zeros and poles are of order one. In Fig.~\ref{fig:PD1}, we show
the topological phase diagram of NHSSH $(0,1)$ chain, its GBZ and
zero/pole configurations for some typical cases. The phase diagram
in Fig.~\ref{fig:PD1}(a) has two topologically distinct gapped phases
with $W=0$ and $1$. They are separated by four phase
boundaries {[}Eq.~(\ref{eq:r0PBaa}){]}, where the two non-Bloch bands
of the system touches with each other. Along the dotted parameter
line in Fig.~\ref{fig:PD1}(a), we pick up five representative points
$\CIRCLE$, $\blacktriangleright$, $\blacklozenge$, $\blacktriangleleft$,
and $\blacksquare$. The GBZ of the system and the zero/pole locations
of the $P(z)$ and $Q(z)$ at these points are shown in Figs.~\ref{fig:PD1}(b)--\ref{fig:PD1}(f).
Our theoretical predictions are clearly verified in each case. For
example, in Fig.~\ref{fig:PD1}(d), we find two zeros (one zero)
of $P(z)$ ($Q(z)$) and one pole of $Q(z)$ inside the GBZ, yielding
the winding numbers $W_{1}=2-0=2$, $W_{2}=1-1=0$, and thus $W=(W_{1}+W_{2})/2=1$.
In Fig.~\ref{fig:PD1}(e), we find only a zero of $P(z)$ and a pole
of $Q(z)$ inside the GBZ, yielding the winding numbers $W_{1}=1-0=1$,
$W_{2}=0-1=-1$, and $W=(W_{1}+W_{2})/2=0$. This critical point is
thus topologically \emph{trivial}. Note in passing that if we only
take into account the contribution of $P(z)$ inside the GBZ, as did
in previous studies for gapped phases \cite{NHSE07}, we will obtain
$W=W_{1}=1$, leading to the prediction that the critical point at
$(J,\gamma)=(\sqrt{2},\sqrt{3})$ is topologically nontrivial.
We will soon verify that this prediction is \textit{incorrect}, as there are
no edge zero modes at this critical point. Our work then generalizes
previous theories on gapped non-Hermitian topological phases to a
rule applicable to both gapped and gapless situations.

\begin{figure}
	\begin{centering}
		\includegraphics[scale=0.48]{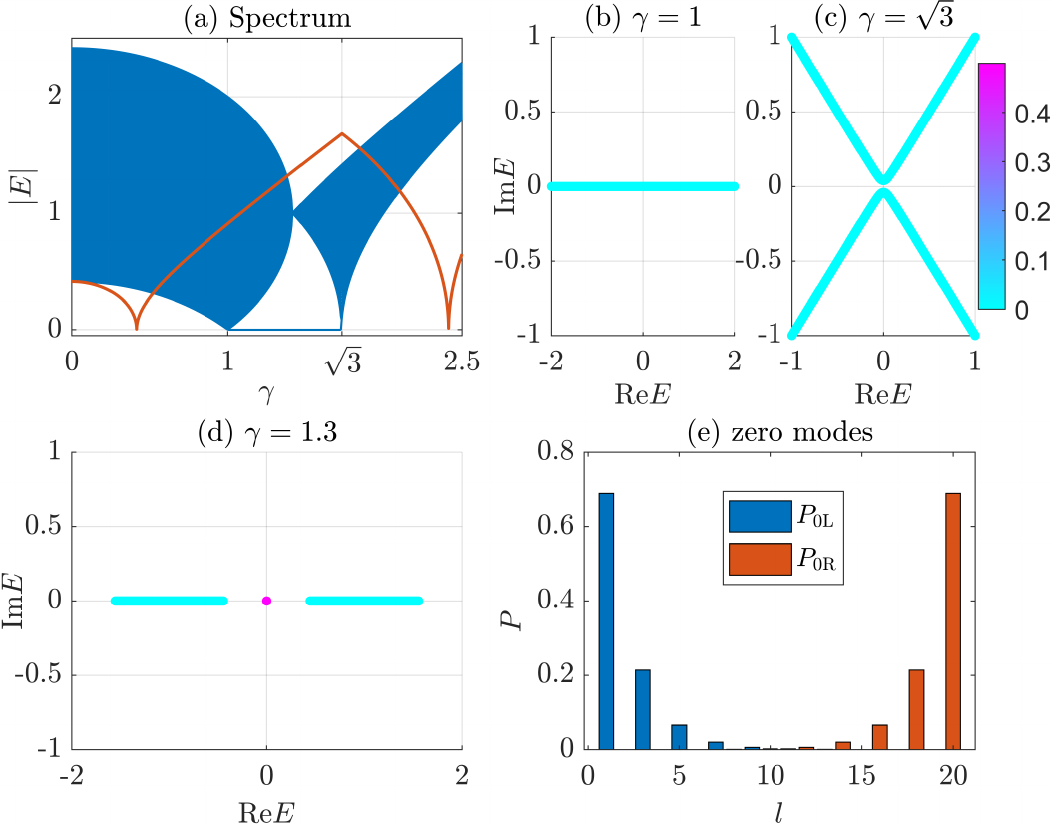}
		\par\end{centering}
	\caption{Spectra and edge states of the NHSSH $(0,1)$ chain with $J=\sqrt{2}$.
		(a) shows the absolute value of spectrum $|E|$ vs $\gamma$ under
		the OBC (in blue dots), with the red curve denoting the magnitude
		of spectral gap at $E=0$ under the PBC. (b)--(d) show the spectra
		at different $\gamma$ on the complex plane, with the shared color
		bar giving the inverse participation ratio of each eigenstate. (e)
		shows the probability distributions $P_{0{\rm L}}$ and $P_{0{\rm R}}$
		of the two zero modes in (d) for a lattice with $10$ unit cells.
		\label{fig:EP1}}
\end{figure}

In Fig.~\ref{fig:EP1}, we present the spectra and edge states of
the NHSSH $(0,1)$ chain. In Fig.~\ref{fig:EP1}(a), the blue dots
correspond to the spectrum obtained under the OBC, whereas the red
line describes the spectrum gap $\min_{k\in[-\pi,\pi]}|E(k)|$ obtained
under the PBC. It is clear that the PBC and OBC spectra have different
gap closing points, which demonstrates the breakdown of Hermitian
bulk-edge correspondence and the necessity of incorporating the non-Bloch
band theory to characterize the NHSSH $(0,1)$ chain. In the parameter
region $\gamma\in(1,\sqrt{3})$, we further observe eigenmodes
pinned at zero energy, which represent topological edge modes. As
illustrated by the OBC spectra in Figs.~\ref{fig:EP1}(b) and \ref{fig:EP1}(c),
these edge modes are absent at the critical points $\gamma=1,\sqrt{3}$,
which confirms the topological triviality of gap closing (phase transition)
points in the NHSSH $(0,1)$ chain. The OBC spectrum at $\gamma=1.3$
in Fig.~\ref{fig:EP1}(d) instead holds a pair of zero modes in the
bulk gap, whose spatial profiles are shown in Fig.~\ref{fig:EP1}(e).
These zero modes are indeed localized at the edges of the lattice,
which verifies the prediction of winding number $W=1$ and number
of edge modes $N_{0}=2$ at $(J,\gamma)=(\sqrt{2},1.3)$ in the phase
diagram Fig.~\ref{fig:PD1}(a). Overall, the results presented in
Fig.~\ref{fig:EP1} are consistent with our theoretical descriptions, which further affirms the applicability
of our approach to the topological characterization of both gapped
and gapless  phases in sublattice-symmetric non-Hermitian
systems.

We next consider the case with $(\alpha,\alpha')=(1,2)$. The complex-extended
Hamiltonian reads $H_{12}(z)\equiv H_{1}(z)+H_{2}(z)$. For the hopping
parameters, we take $J_{1}^{{\rm L}}=J-\gamma$, $J_{1}^{{\rm R}}=J+\gamma$
($J,\gamma\in\mathbb{R}$), and let $J_{2}^{{\rm L}}=J_{2}^{{\rm R}}=1$
be the unit of energy. The $H_{12}(z)$ is also non-Hermitian when
$\gamma\neq0$. The non-Hermiticity is now due to asymmetric
intercell hoppings $J_{1}^{{\rm L,R}}$. Following Eqs.~(\ref{eq:GBZaa})
and (\ref{eq:PBaa}), the GBZ radius $r_{0}$ and phase boundary equation
in this case are still given by the Eq.~(\ref{eq:r0PBaa}). While
the topological invariant $W$ and the number of zero-energy
edge modes $N_{0}$ are predicted by the Eqs.~(\ref{eq:Waa}) and
(\ref{eq:BBCaa}) as
\begin{alignat}{1}
	W&=\begin{cases}
		2, & |J^{2}-\gamma^{2}|<1\\
		1, & {\rm Otherwise}
	\end{cases},\label{eq:Waa2}\\
	N_{0}&=\begin{cases}
	4, & |J^{2}-\gamma^{2}|<1\\
	2, & {\rm Otherwise}
	\end{cases}.\label{eq:BBCaa2}
\end{alignat}
We notice that even though the GBZ radii and phase boundaries of $H_{01}(z)$
and $H_{12}(z)$ are identical, their winding numbers and edge mode
configurations are different in each phase. Importantly, the critical
points of $H_{12}(z)$ along the phase boundary $J^{2}=\gamma^{2}\pm1$
are all topologically nontrivial, characterized by the invariant
$W=1$ and the number of edge zero modes $N_{0}=2$. Despite the point
at $\gamma=0$, these topologically nontrivial critical points are
unique to non-Hermitian systems, as the conventional bulk-edge correspondence
breaks down at each of these points due to the difference between
the BZ and GBZ.

The Eqs.~(\ref{eq:Waa2}) and (\ref{eq:BBCaa2}) can
also be verified by computing the zero-energy solutions of the system.
In the lattice representation, the Hamiltonian of NHSSH $(1,2)$ chain
takes the form $\hat{H}_{12}=\hat{H}_{1}+\hat{H}_{2}$, where $\hat{H}_{1}=\sum_{n}[(J-\gamma)\hat{b}_{n}^{\dagger}\hat{a}_{n+1}+(J+\gamma)\hat{a}_{n+1}^{\dagger}\hat{b}_{n}]$
and $\hat{H}_{2}=\sum_{n}(\hat{b}_{n}^{\dagger}\hat{a}_{n+2}+\hat{a}_{n+2}^{\dagger}\hat{b}_{n})$.
For a chain with unit cell indices $n=1,2,...,N$ under the OBC, the (non-normalized) zero-energy eigenstates are found
in the limit $N\rightarrow\infty$ and in symmetrized basis as (see Appendix \ref{sec:App1}
for details)
\begin{alignat}{1}
	|\varphi_{0}^{{\rm L,1}}\rangle= & \hat{a}_{1}^{\dagger}|\emptyset\rangle,\qquad|\varphi_{0}^{{\rm R,1}}\rangle= \hat{b}_{N}^{\dagger}|\emptyset\rangle,\nonumber\\
	|\varphi_{0}^{{\rm L,2}}\rangle= & \sum_{n=2}^{N}[-\varrho(J-\gamma)]^{n-2}\hat{a}_{n}^{\dagger}|\emptyset\rangle,\nonumber \\
	|\varphi_{0}^{{\rm R,2}}\rangle= & \sum_{n=1}^{N-1}[-\varrho^{-1}(J+\gamma)]^{N-n-1}\hat{b}_{n}^{\dagger}|\emptyset\rangle,\label{eq:H12Edge}
\end{alignat}
where $|\varrho|=\sqrt{|J+\gamma|/|J-\gamma|}$. The eigenmodes $|\varphi_{0}^{{\rm L,1}}\rangle$
and $|\varphi_{0}^{{\rm R,1}}\rangle$ persist at the two ends of
the chain throughout the parameter space. They could
thus survive at the critical points $J^{2}=\gamma^{2}\pm1$, making
the latter topologically nontrivial with $W=1$
and the number of edge zero modes $N_{0}=2$. Meanwhile, the states $|\varphi_{0}^{{\rm L,2}}\rangle$
and $|\varphi_{0}^{{\rm R,2}}\rangle$ form a pair of edge
zero modes if and only if $|J^{2}-\gamma^{2}|<1$ (see Appendix \ref{sec:App1}
for details). Therefore, they could only survive in the gapped phase
of the chain with $W=2$ and the number of edge zero modes
$N_{0}=4$, as described by the Eqs.~(\ref{eq:Waa2}) and (\ref{eq:BBCaa2}).
In the region $|J^{2}-\gamma^{2}|>1$, we are left with
the edge zero modes $|\varphi_{0}^{{\rm L,1}}\rangle$ and $|\varphi_{0}^{{\rm R,1}}\rangle$,
which is coincident with the gapped topological phase of $W=1$ and
$N_{0}=2$ according to the Eqs.~(\ref{eq:Waa2}) and (\ref{eq:BBCaa2}).
These observations confirm that our theory is valid in both the gapped
topological phases and along the gapless phase boundaries of the NHSSH
$(1,2)$ chain. Importantly, the nontrivial critical points of non-Bloch
bands are correctly depicted within our topological characterization.

\begin{figure}
	\begin{centering}
		\includegraphics[scale=0.5]{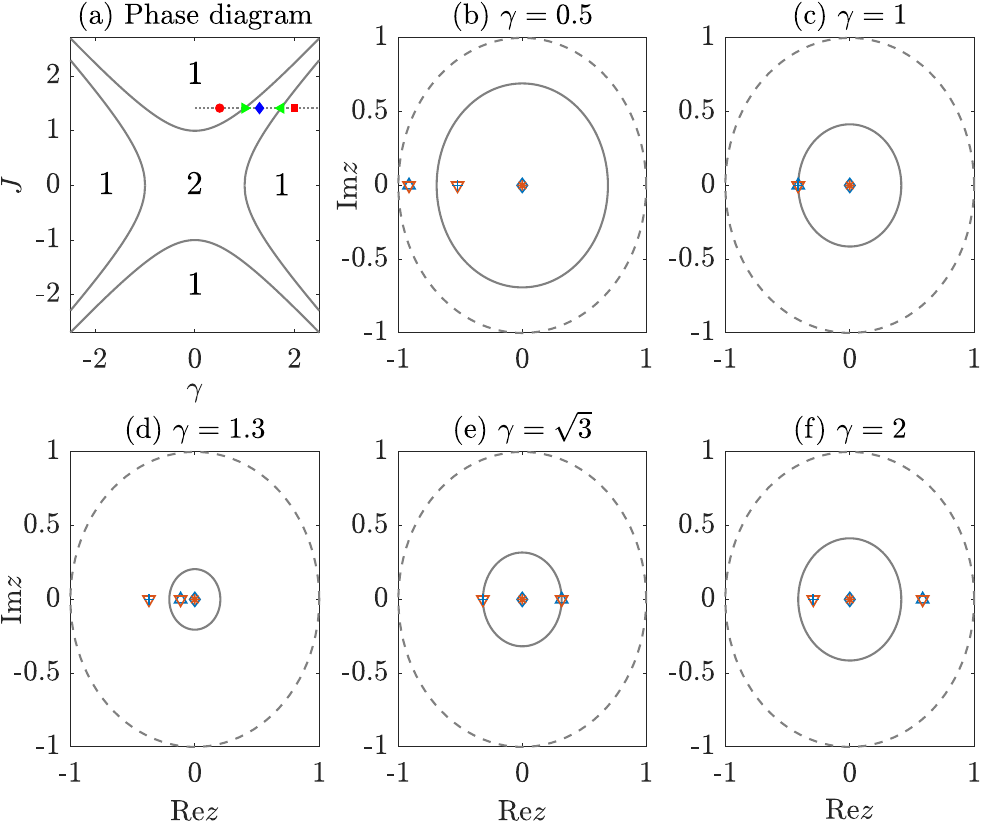}
		\par\end{centering}
	\caption{Phase diagram, GBZ and zero/pole distributions of the NHSSH $(1,2)$
		chain. (a) Topological phase diagram, with the values of
		$W$ shown explicitly in each phase. Phase boundaries are given by
		the gray solid lines. Along the dotted line, the symbols $\CIRCLE$,
		$\blacktriangleright$, $\blacklozenge$, $\blacktriangleleft$, and
		$\blacksquare$ highlight the points in the parameter space with $J=\sqrt{2}$
		and $\gamma=0.5,1,1.3,\sqrt{3}$, and $2$, respectively. (b)--(f)
		show the BZ (in dashed rings), GBZ (in solid rings), zeros
		of $P(z)$ (in $\lozenge$ and $\vartriangle$ for the third order
		and first order ones), poles of $P(z)$ (in $+$), zeros of $Q(z)$
		(in $\triangledown$), and poles of $Q(z)$ (in $*$) for the five
		exemplary points in (a) from left to right on the complex $z$-plane. \label{fig:PD2}}
\end{figure}

In the rest of this subsection, we offer additional evidence to support
our findings. Inserting $(J_{1}^{{\rm L}},J_{1}^{{\rm R}})=(J-\gamma,J+\gamma)$
and $(J_{2}^{{\rm L}},J_{2}^{{\rm R}})=(1,1)$ into the Eqs.~(\ref{eq:Paaz})
and (\ref{eq:Qaaz}), we realize that the $P(z)$ has two zeros at
$z=0$ (order $3$), $z=-(J-\gamma)$ (order $1$), and a pole at
$z=-1/(J+\gamma)$ (order $1$). The $Q(z)$ has two zeros at $z=-(J-\gamma),-1/(J+\gamma)$
and a pole at $z=0$. The zeros and poles of $Q(z)$ are all of order
one. In Fig.~\ref{fig:PD2}, we show the topological phase diagram
of our NHSSH $(1,2)$ chain, its GBZ and zero/pole locations for typical
cases. We notice that the phase boundaries in Figs.~\ref{fig:PD1}(a)
and \ref{fig:PD2}(a) have the same configurations, as described by
the Eq.~(\ref{eq:r0PBaa}). Nevertheless, the invariant $W$
of their corresponding phases are different, and the gapped phases
of NHSSH $(1,2)$ chain are all topologically nontrivial. In Fig.~\ref{fig:PD2}(a), we take the same representative points at $\CIRCLE$,
$\blacktriangleright$, $\blacklozenge$, $\blacktriangleleft$, and
$\blacksquare$ as in Fig.~\ref{fig:PD1}(a), and show their respective
GBZ and zero/pole distributions of $P(z)$ and $Q(z)$ in Fig.~\ref{fig:PD2}(b)--(f).
The results are again consistent with our predictions in each case.
Different from the $(0,1)$ chain, the critical points of the $(1,2)$
chain at $(J,\gamma)=(\sqrt{2},1)$ {[}Fig.~\ref{fig:PD2}(c){]} and
$(J,\gamma)=(\sqrt{2},\sqrt{3})$ {[}Fig.~\ref{fig:PD2}(e){]} (and
along other phase boundaries) are found to be topologically nontrivial.
For example, at $(J,\gamma)=(\sqrt{2},\sqrt{3})$ in Fig.~\ref{fig:PD2}(a),
we have $W_{1}=3-0=3$ and $W_{2}=0-1=-1$ for the $P(z)$ and $Q(z)$
in Fig.~\ref{fig:PD2}(e), yielding the invariant $W=(W_{1}+W_{2})/2=1$.
We will demonstrate that there is indeed a number of $N_{0}=2|W|=2$
edge zero modes at this critical point, verifying our prediction in
Eq.~(\ref{eq:BBCaa2}). It deserves to emphasize again that taking
only the zero/pole contribution of $P(z)$ into account will
lead to $W=W_{1}=3$, which could not generate any reliable predictions
about the numbers of edge zero modes that could appear at the critical
point.

\begin{figure}
	\begin{centering}
		\includegraphics[scale=0.48]{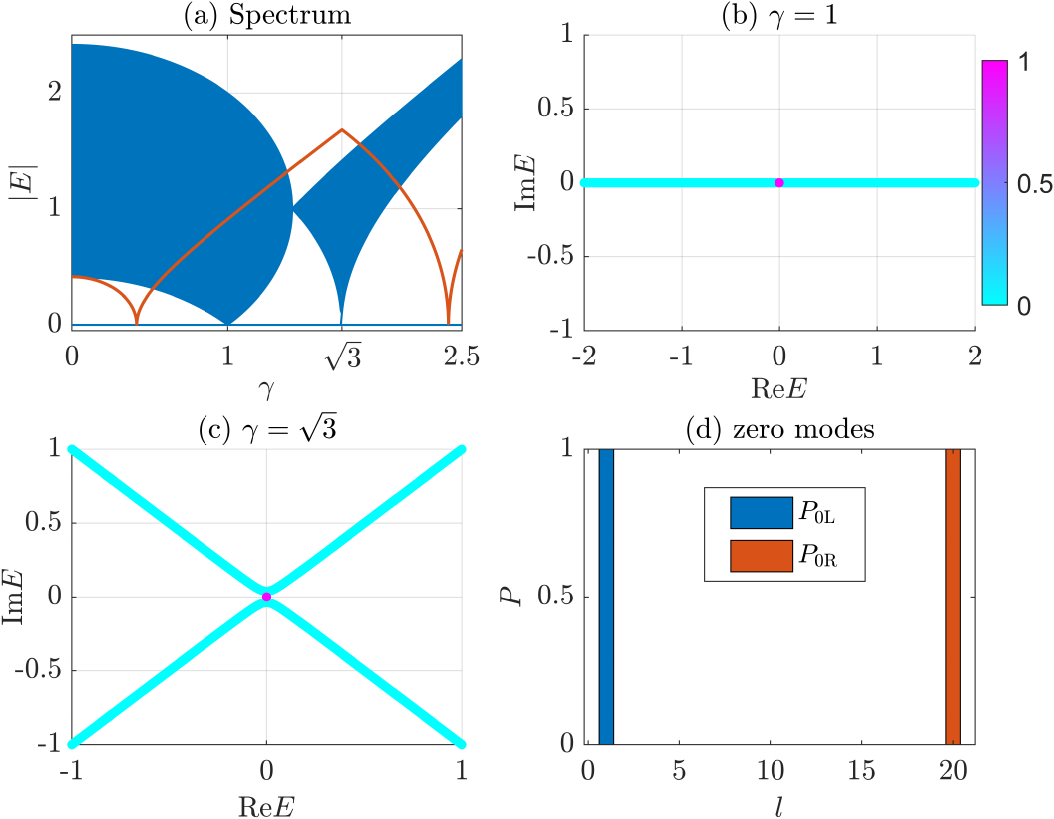}
		\par\end{centering}
	\caption{Spectra and edge states of the NHSSH $(1,2)$ chain with $J=\sqrt{2}$.
		(a) shows the absolute value of spectrum $|E|$ vs $\gamma$ under
		the OBC (in blue dots), with the red curve denoting the magnitude
		of spectral gap at $E=0$ under the PBC. (b) and (c) show the spectra
		at $\gamma=1$ and $\sqrt{3}$ on the complex plane, with the shared
		color bar giving the inverse participation ratio of each eigenstate.
		(d) shows the probability distributions $P_{0{\rm L}}$ and $P_{0{\rm R}}$
		of the two zero modes in (b) for a lattice with $10$ unit cells.
		The two edge modes in (c) have the same profiles $P_{0{\rm L}}$ and
		$P_{0{\rm R}}$. \label{fig:EP2}}
\end{figure}

In Fig.~\ref{fig:EP2}, we present the spectra and edge states of
the NHSSH $(1,2)$ chain. We observe that under the OBC, zero energy
eigenmodes persist throughout the spectrum (blue dots) in Fig.~\ref{fig:EP2}(a).
Meanwhile, the gap closing points under the PBC and OBC are again
different, making it necessary to employ the non-Bloch band theory
for topological characterizations. Notably, we observe strongly localized
zero modes in the spectra at the critical points $(J,\gamma)=(\sqrt{2},1)$
and $(\sqrt{2},\sqrt{3})$, as presented in Figs.~\ref{fig:EP2}(b)
and \ref{fig:EP2}(c). It implies that in stark contrast to the NHSSH
$(0,1)$ chain, the band touching points of NHSSH $(1,2)$ chain
are topologically nontrivial. Indeed, the zero modes in
Figs.~\ref{fig:EP2}(b) and \ref{fig:EP2}(c) are both localized at
the two ends of the chain, forming a pair of degenerate edge modes
at $E=0$ as shown in Fig.~\ref{fig:EP2}(d). On the other hand, our
bulk theory yields the winding number $W=1$ at both critical points,
leading to the consistent bulk-edge correspondence $N_{0}=2|W|$.
Therefore, beyond previous approaches, our theory allows us to identify
a unique class of topologically nontrivial gapless points between
non-Bloch bands and characterize the associated bulk-edge correspondence
in 1D non-Hermitian systems.

\begin{figure}
	\begin{centering}
		\includegraphics[scale=0.48]{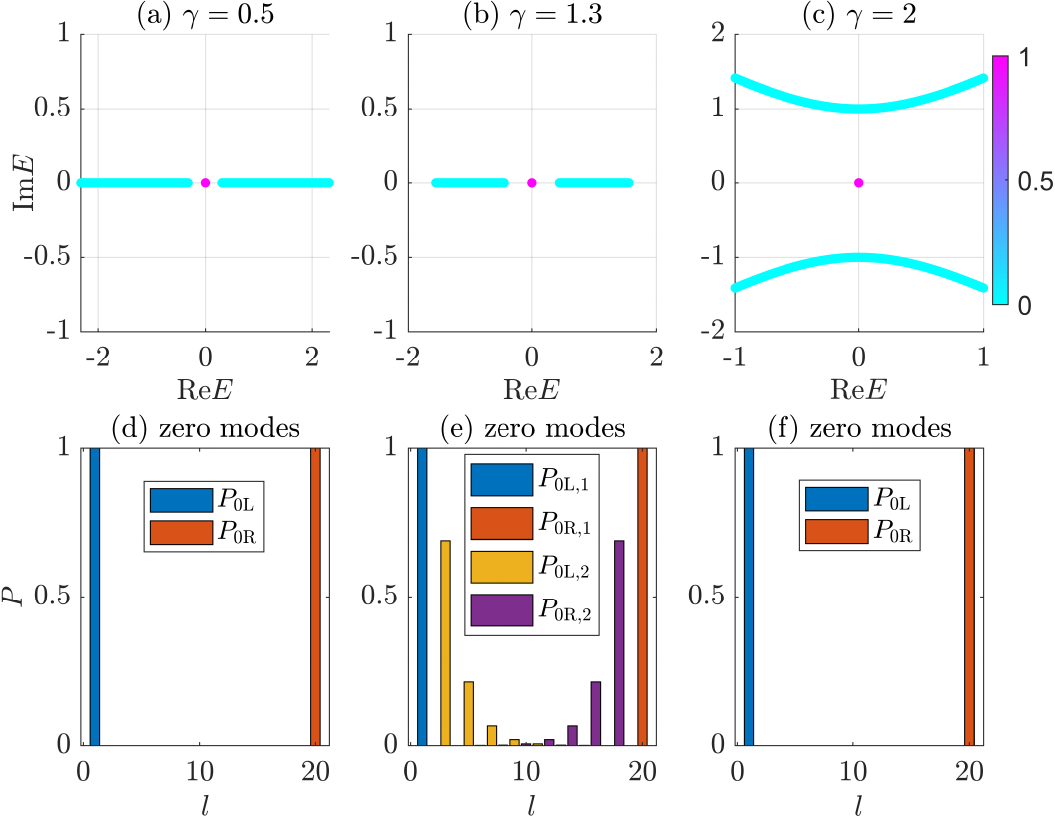}
		\par\end{centering}
	\caption{Spectra and edge states of the NHSSH $(1,2)$ chain with $J=\sqrt{2}$
		for three gapped phases. In (a), there are two edge zero modes, whose
		probability distributions $P_{0{\rm L}}$ and $P_{0{\rm R}}$ are
		shown in (d) for a lattice with $10$ unit cells. In (b), there are
		four edge zero modes, whose probability distributions $P_{0{\rm L},1}$,
		$P_{0{\rm R},1}$, $P_{0{\rm L},2}$ and $P_{0{\rm R},2}$ are presented
		in (e) for a lattice with $10$ unit cells. In (c), there are two
		edge zero modes, whose probability distributions $P_{0{\rm L}}$
		and $P_{0{\rm R}}$ are given in (f) for a lattice with $10$ unit
		cells. \label{fig:GAP2}}
\end{figure}

For completeness, we present the spectra and edge states of the NHSSH
$(1,2)$ chain for three gapped phases in Fig.~\ref{fig:GAP2}, whose
system parameters are given by $(J,\gamma)=(\sqrt{2},0.5)$,
$(\sqrt{2},1.3)$ and $(\sqrt{2},2)$ in the phase diagram Fig.~\ref{fig:PD2}(a).
According to the phase diagram, the topological invariants of these
phases are $W=1,2,1$, while the numbers of edge zero modes
in Figs.~\ref{fig:GAP2}(d)--\ref{fig:GAP2}(f) are $N_{0}=2,4,2$.
The bulk-edge correspondence $N_{0}=2|W|$ is thus verified for these
gapped phases. In conclusion, we have demonstrated the applicability
of our theory to the topological characterization of NHSSH $(\alpha,\alpha')$
chains for both gapped and gapless phases. In the next subsection,
we study a more complicated class of NHSSH chain and reveal the presence
of phase transitions along topological phase boundaries.

\subsection{$(\alpha,\alpha',\alpha'')$ chain: critical topological phase transitions\label{subsec:SSHaaa}}

As a final example, we consider a NHSSH chain with three
hopping ranges. Extending $e^{ik}$ to the whole complex plane, the
Hamiltonian of such a NHSSH $(\alpha,\alpha',\alpha'')$ chain reads
$H(z)=H_{\alpha}(z)+H_{\alpha'}(z)+H_{\alpha''}(z)$ where $\alpha\neq\alpha'\neq\alpha''$.
To simplify the analytical treatments and pinpoint the key physics,
we focus on the case with $\alpha\geq0$, $\alpha'=\alpha+1$ and
$\alpha''=\alpha+2$. Under these assumptions, we find the characteristic
functions $P(z)$ and $Q(z)$ of NHSSH $(\alpha,\alpha+1,\alpha+2)$
chain as
\begin{alignat}{1}
	P(z)= & z^{2\alpha+2}\frac{J_{\alpha+2}^{{\rm L}}z^{2}+J_{\alpha+1}^{{\rm L}}z+J_{\alpha}^{{\rm L}}}{J_{\alpha}^{{\rm R}}z^{2}+J_{\alpha+1}^{{\rm R}}z+J_{\alpha+2}^{{\rm R}}},\label{eq:Paaa}\\
	Q(z)= & \frac{(J_{\alpha+2}^{{\rm L}}z^{2}+J_{\alpha+1}^{{\rm L}}z+J_{\alpha}^{{\rm L}})(J_{\alpha}^{{\rm R}}z^{2}+J_{\alpha+1}^{{\rm R}}z+J_{\alpha+2}^{{\rm R}})}{z^{2}}.\label{eq:Qaaa}
\end{alignat}
It can be verified that the $P(z)$ has a zero of order $2\alpha+2$
at $z=0$, two other zeros of order one at
\begin{equation}
	z_{\pm}^{{\rm L}}=\frac{-J_{\alpha+1}^{{\rm L}}\pm\sqrt{(J_{\alpha+1}^{{\rm L}})^{2}-4J_{\alpha}^{{\rm L}}J_{\alpha+2}^{{\rm L}}}}{2J_{\alpha+2}^{{\rm L}}},\label{eq:zpmL}
\end{equation}
and two poles of order one at
\begin{equation}
	z_{\pm}^{{\rm R}}=\frac{-J_{\alpha+1}^{{\rm R}}\pm\sqrt{(J_{\alpha+1}^{{\rm R}})^{2}-4J_{\alpha}^{{\rm R}}J_{\alpha+2}^{{\rm R}}}}{2J_{\alpha}^{{\rm R}}}.\label{eq:zpmR}
\end{equation}
Meanwhile, the $Q(z)$ has for zeros of order one at $z_{\pm}^{{\rm L,R}}$
and a pole of order two at $z=0$. The locations of these zeros and
poles vs the GBZ determine the topological invariants
of the system following Eqs.~(\ref{eq:W12}) and (\ref{eq:W}).

To proceed, we need to obtain the GBZ from the two middle solutions
of the equation $Q(z)=Q(ze^{i\theta})$ for $\theta\in[0,2\pi]$ \cite{NHSE03}.
As a further simplification, we take $J_{\beta}^{{\rm L}}=J_{\beta}^{{\rm R}}=J_{\beta}\in\mathbb{C}$
with $\beta=\alpha,\alpha+1,\alpha+2$ for the hopping amplitudes
and set $J_{\alpha+1}=1$ as the unit of energy. The system Hamiltonian
remains non-Hermitian whenever there exists a $\beta\in\{\alpha,\alpha+1,\alpha+2\}$
such that $J_{\beta}\neq J_{\beta}^{*}$. Substituting these conditions
into the equation $Q(z)=Q(ze^{i\theta})$, we obtain $(J_{\alpha+2}+J_{\alpha})ze^{i\theta}+J_{\alpha}J_{\alpha+2}(e^{i\theta}+1)=J_{\alpha}J_{\alpha+2}z^{4}e^{i2\theta}(e^{i\theta}+1)+(J_{\alpha+2}+J_{\alpha})z^{3}e^{i2\theta}$.
For each given set of parameters $(J_{\alpha},J_{\alpha+2},\theta)$,
this equation has four solutions in $z$, with two of them
given by $z_{\pm}=\pm e^{-i\theta/2}$. As these solutions have equal
magnitudes $|z_{+}|=|z_{-}|=1$, they must be the two middle solutions
that determine the GBZ of the system. Therefore, the GBZ radius is
given by $r_{0}=|z_{\pm}|=1$, which means that it is identical to
the standard BZ. Nevertheless, we will show that there are still topological
phases and transitions induced by non-Hermitian effects, which are
captured by our theory.

\begin{figure}
	\begin{centering}
		\includegraphics[scale=0.49]{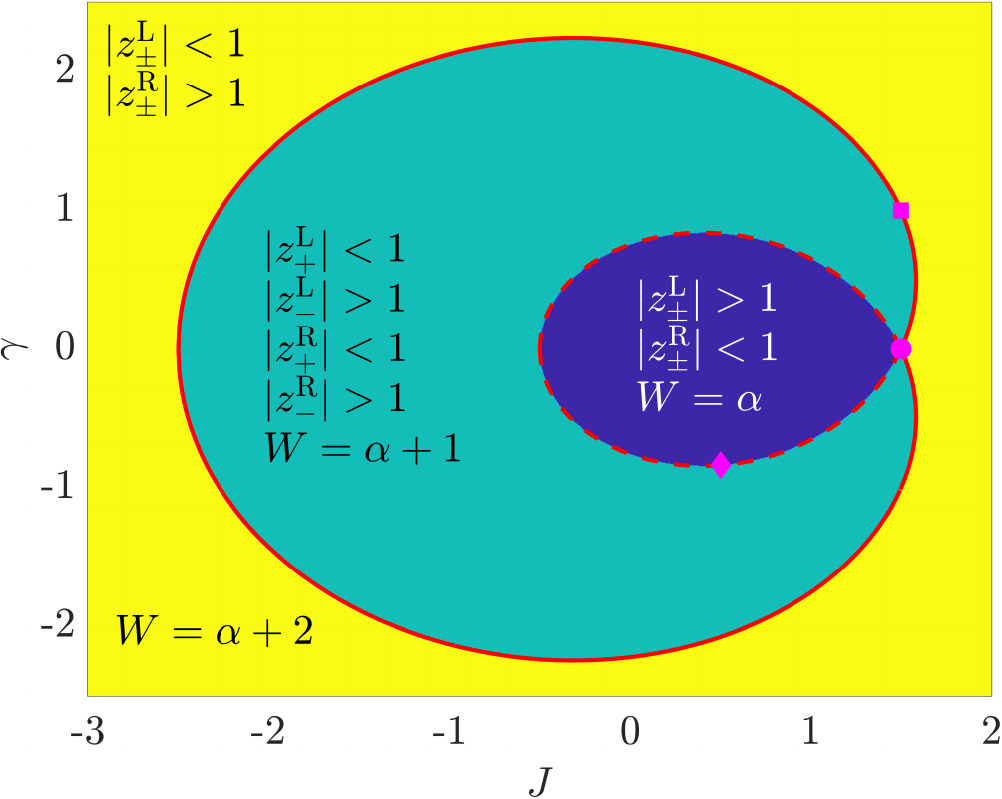}
		\par\end{centering}
	\caption{Topological phase diagram of the NHSSH $(\alpha,\alpha+1,\alpha+2)$
		chain. System parameters are $J_{\alpha}^{{\rm L}}=J_{\alpha}^{{\rm R}}=J_{\alpha}=1.5$,
		$J_{\alpha+1}^{{\rm L}}=J_{\alpha+1}^{{\rm R}}=J_{\alpha+1}=1$ and
		$J_{\alpha+2}^{{\rm L}}=J_{\alpha+2}^{{\rm R}}=J_{\alpha+2}=J+i\gamma$,
		where $J,\gamma\in\mathbb{R}$. Each region with a uniform color refers
		to a gapped phase, with its topological invariant $W$ and zero/pole
		locations $|z_{\pm}^{{\rm L}}|,|z_{\pm}^{{\rm R}}|$ relative to GBZ
		shown explicitly therein. The red dashed and solids lines denote
		two sets of topologically distinct phase boundaries with $W=\alpha$
		and $\alpha+1$. The magenta $\blacklozenge$, $\CIRCLE$ and $\blacksquare$ highlight three critical points. \label{fig:PD012}}
\end{figure}

In Fig.~\ref{fig:PD012}, we show the topological phase diagram of
NHSSH $(\alpha,\alpha+1,\alpha+2)$ chain for a representative case.
Each region with a uniform color corresponds to a gapped
phase, while the red dashed and dotted lines depict topological phase
boundaries. In the region surrounded by the red dashed lines, we have
$|z_{\pm}^{{\rm L}}|>1$ and $|z_{\pm}^{{\rm R}}|<1$. The winding
numbers of $P(z)$ and $Q(z)$ are thus given by $W_{1}=2\alpha$
and $W_{2}=0$, yielding the topological invariant $W=\alpha$ for
this phase. In the region sandwiched between the red
dashed and solid lines, we have $|z_{+}^{{\rm L}}|<1$, $|z_{-}^{{\rm L}}|>1$,
$|z_{+}^{{\rm R}}|<1$ and $|z_{-}^{{\rm R}}|>1$. The winding numbers
of $P(z)$ and $Q(z)$ are thus $W_{1}=2\alpha+2$ and $W_{2}=0$,
leading to the topological invariant $W=\alpha+1$ for this phase.
In the region outside the red solid lines, we have $|z_{\pm}^{{\rm L}}|<1$
and $|z_{\pm}^{{\rm R}}|>1$. The winding numbers of $P(z)$ and $Q(z)$
are then given by $W_{1}=2\alpha+4$ and $W_{2}=0$, generating the
topological invariant $W=\alpha+2$ for this phase. The
red dashed line in Fig.~\ref{fig:PD012} is given by the solution
of $|z_{+}^{{\rm L}}|=|z_{-}^{{\rm R}}|=1$. On this dashed line,
we have $|z_{-}^{{\rm L}}|>1$ and $|z_{+}^{{\rm R}}|<1$. Therefore,
along the phase boundary depicted by the red dashed line, we find
the winding numbers of $P(z)$ and $Q(z)$ as $W_{1}=2\alpha+1$ and
$W_{2}=-1$, yielding the topological invariant $W=\alpha$. The red dashed critical line in Fig.~\ref{fig:PD012} is thus topologically
trivial (nontrivial) if $\alpha=0$ ($\alpha>0$). On the other hand,
the red solid line in Fig.~\ref{fig:PD012} is given by the solution
of $|z_{-}^{{\rm L}}|=|z_{+}^{{\rm R}}|=1$. Along this line, we
find $|z_{+}^{{\rm L}}|<1$ and $|z_{-}^{{\rm R}}|>1$. Therefore,
on the phase boundary depicted by the red solid line,
the winding numbers of $P(z)$ and $Q(z)$ are $W_{1}=2\alpha+3$ and
$W_{2}=-1$, yielding the topological invariant $W=\alpha+1$. The red solid critical line in Fig.~\ref{fig:PD012}
is thus topologically nontrivial for all $\alpha\geq0$. 

Up to now, we have achieved the topological characterization of all
gapped phases and gapless phase boundaries for our considered NHSSH
$(\alpha,\alpha+1,\alpha+2)$ chain. Two additional points deserve
to be mentioned. First, by increasing the strength of non-Hermitian
parameter $\gamma$, we could obtain phases with enlarged topological
invariants ($W=\alpha\rightarrow\alpha+1\rightarrow\alpha+2$) and
thus enriched topological signatures. These non-Hermiticity induced
topological properties are fully captured by our theory.
Second, by changing the system parameter $J_{\alpha+2}=J+i\gamma$
across the multicritical point $(J,\gamma)=(1.5,0)$ smoothly from
the red dashed to solid lines, we will encounter a topological phase
transition with the $W$ going from $\alpha$ to $\alpha+1$.
This transition does not require the closing and reopening of any
bulk spectral gaps at zero energy. Therefore, we could have a phase
transition along topologically distinct gapless phase boundaries in
the NHSSH $(\alpha,\alpha+1,\alpha+2)$ chain. The topological character
of such a transition is again well described by our approach, while
existing theories developed for gapped non-Hermitian phases (either
with point or line gaps) fail to do so.

\begin{figure}
	\begin{centering}
		\includegraphics[scale=0.48]{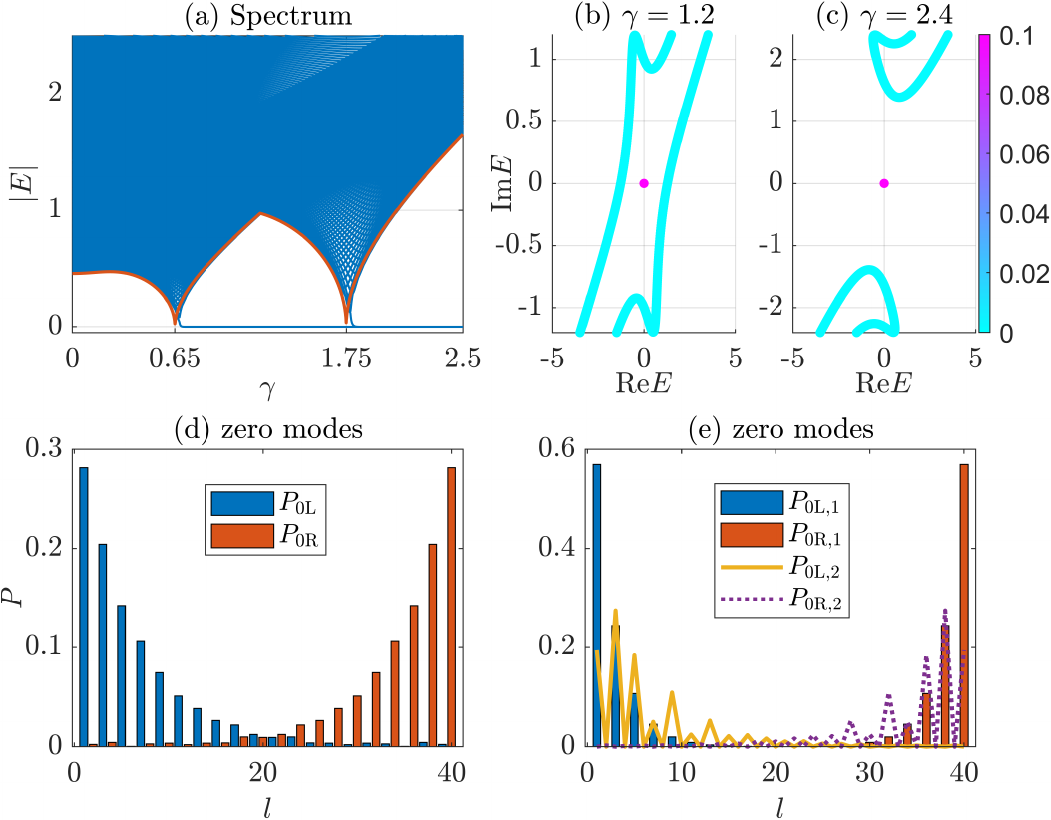}
		\par\end{centering}
	\caption{Spectra and edge states of the NHSSH $(\alpha,\alpha+1,\alpha+2)$
		chain with $\alpha=0,$ $J_{\alpha}^{{\rm L}}=J_{\alpha}^{{\rm R}}=J_{\alpha}=1.5$,
		$J_{\alpha+1}^{{\rm L}}=J_{\alpha+1}^{{\rm R}}=J_{\alpha+1}=1$, $J_{\alpha+2}^{{\rm L}}=J_{\alpha+2}^{{\rm R}}=J_{\alpha+2}=J+i\gamma$
		and $J=1$. (a) shows the absolute value of spectrum $|E|$ vs $\gamma$
		under the OBC (in blue dots), with the red curve denoting the magnitude
		of spectral gap at $E=0$ under the PBC. (b) and (c) show the spectra
		at $\gamma=1.2$ and $2.4$ on the complex plane, with the shared
		color bar giving the inverse participation ratio of each eigenstate.
		(d) shows the probability distributions $P_{0{\rm L}}$ and $P_{0{\rm R}}$
		of the two zero modes in (b) for a lattice with $20$ unit cells.
		(e) shows the probability distributions $P_{0{\rm L},1}$, $P_{0{\rm R},1}$,
		$P_{0{\rm L},2}$ and $P_{0{\rm R},2}$ of the four zero modes in (c)
		for a lattice with $20$ unit cells. \label{fig:EP3}}
\end{figure}

We next illustrate our results about the NHSSH $(\alpha,\alpha+1,\alpha+2)$
chain with numerical examples. To be explicit, we focus on the case
with $\alpha=0$, set the second neighbor hopping $J_{2}=J+i\gamma$
as the complex parameter, and fix other system parameters as $(J_{0},J_{1},J)=(1.5,1,1)$.
The resulting energy spectrum is shown in Fig.~\ref{fig:EP3}(a).
With the increase of $\gamma$, we observe two phase transitions accompanied
by the closing/reopening of bulk spectral gaps (highlighted by the
red solid line) at $\gamma\approx0.65$ and $\gamma\approx1.75$.
After each transition, more eigenmodes appear in the spectrum under
the OBC (blue dots) at zero energy. These zero modes are spatially
localized, as exemplified by their inverse participation ratios in
the spectra reported in Figs.~\ref{fig:EP3}(b) and \ref{fig:EP3}(c)
for $\gamma=1.2$ and $2.4$, respectively. These zero modes are also
found to stay at the two edges of the system, as shown in Figs.~\ref{fig:EP3}(d)
and \ref{fig:EP3}(e). Further counting reveals that with the raise
of $\gamma$, the number of edge zero modes we obtained following
the two sequential topological transitions are $N_{0}=2$ and $4$.
Meanwhile, according to the phase diagram Fig.~\ref{fig:PD012}, the
three gapped phases in Fig.~\ref{fig:EP3}(a) from left to right have
the topological invariants $W=0$, $1$ and $2$, successively. The
bulk-edge correspondence $N_{0}=2|W|$, as predicted by our theory,
is thus verified for all gapped phases of the NHSSH $(\alpha,\alpha+1,\alpha+2)$
chain with $\alpha=0$. We have also checked and confirmed that other
choices of $\alpha>0$ generate consistent results.

\begin{figure}
	\begin{centering}
		\includegraphics[scale=0.47]{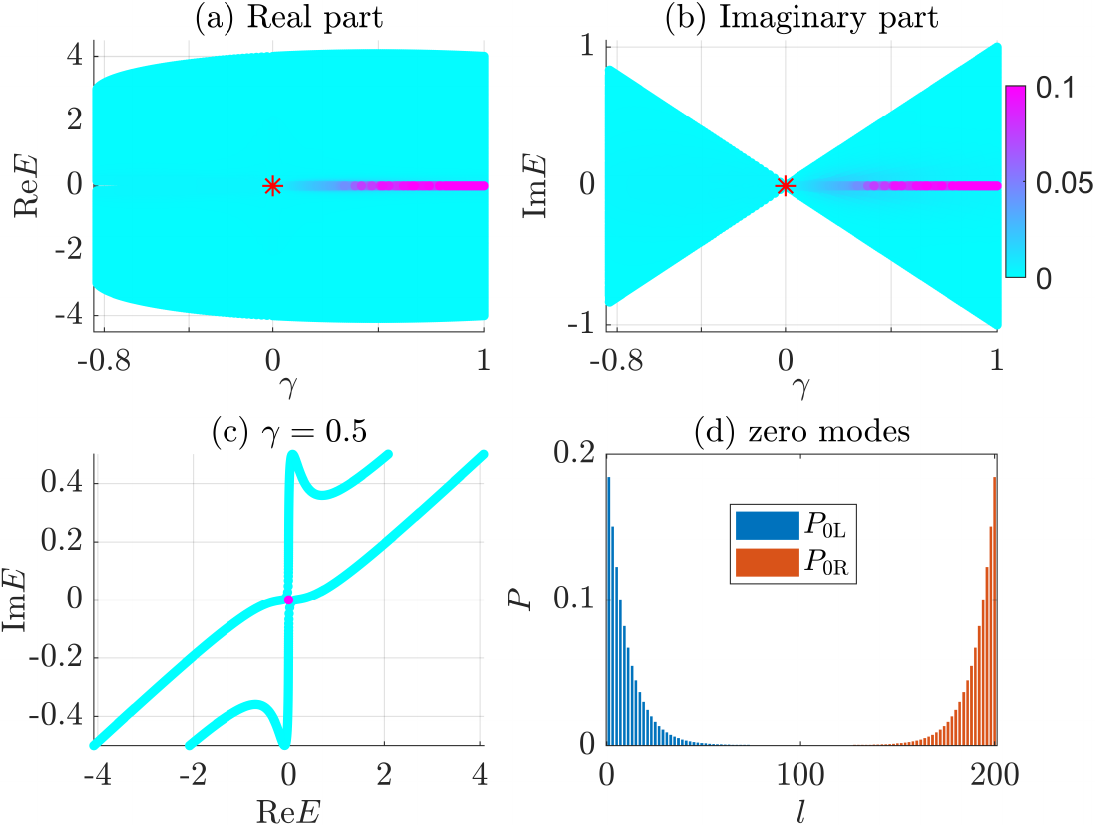}
		\par\end{centering}
	\caption{Spectra and edge states of the NHSSH $(\alpha,\alpha+1,\alpha+2)$
		chain along the critical line. We set $\alpha=0,$ $J_{\alpha}^{{\rm L}}=J_{\alpha}^{{\rm R}}=J_{\alpha}=1.5$,
		$J_{\alpha+1}^{{\rm L}}=J_{\alpha+1}^{{\rm R}}=J_{\alpha+1}=1$, and
		$J_{\alpha+2}^{{\rm L}}=J_{\alpha+2}^{{\rm R}}=J_{\alpha+2}=J+i\gamma$.
		In (a) and (b), the values of $J$ and $\gamma$ are taken along the
		red dashed (solid) line from $\blacklozenge\rightarrow\CIRCLE$ ($\CIRCLE\rightarrow\blacksquare$) in Fig.~\ref{fig:PD012} with $J\in[0.5,1.5]$
		($\gamma\in(0,1]$) for $\gamma\leq0$ ($\gamma>0$). (a)
		and (b) show the real and imaginary parts of the spectrum under the
		OBC, where the red stars highlight the location of multicritical point $\CIRCLE$
		at $(J,\gamma)=(1.5,0)$, and the color of each data point gives the
		inverse participation ratio of the corresponding state. (c) shows
		the spectrum at $\gamma=0.5$ in (a) and (b) on the complex plane.
		(a)--(c) share the same color bar. (d) shows the probability distributions
		of the two critical edge modes at $E=0$ in (c) for a lattice with
		$100$ unit cells. \label{fig:EP4}}
\end{figure}

Finally, we demonstrate the emergence of critical edge states following
phase transitions along non-Hermitian topological phase boundaries
in Fig.~\ref{fig:EP4}. We focus on a parameter path in Fig.~\ref{fig:PD012}
along the dashed critical line with $J\in[0.5,1.5]$ for $\gamma\leq0$ from $\blacklozenge\rightarrow\CIRCLE$,
followed by the solid critical line with $\gamma\in(0,1]$ for $\gamma>0$ from $\CIRCLE\rightarrow\blacksquare$.
These two critical lines, which have been identified as topologically
distinct according to our theory, are interconnected at the multicritical
point $(J,\gamma)=(1.5,0)$ in Fig.~\ref{fig:PD012}. Interestingly,
we observe that by passing through the multicritical point from the
dashed to solid phase boundaries, localized eigenmodes are generated
at zero energy without undergoing any gap closing and reopening processes
in the complex spectrum, as presented in Figs.~\ref{fig:EP4}(a) and
\ref{fig:EP4}(b). Moreover, the emerging zero modes are indeed coexistent
with a gapless bulk and localized spatially around the two ends of
the lattice, as shown in Figs.~\ref{fig:EP4}(c) and \ref{fig:EP4}(d).
These zero modes are thus referred to as non-Hermitian critical edge
states, whose topological origin following the ``\emph{phase transition
	of phase transition}'' along the gapless critical line is well described
by our theory. Indeed, the phase diagram reported in Fig.~\ref{fig:PD012}
predicts $W=0$ and $W=1$ along the dashed and solid phase boundaries.
The bulk-edge correspondence $N_{0}=2|W|$ is then verified not only
in gapped phases, but also along all gapless phase boundaries of the
NHSSH $(\alpha,\alpha+1,\alpha+2)$ chain. Following our definition
of topological invariant $W$, we could now arrive at a consistent
description of topological phases and transitions in 1D, two-band,
sublattice-symmetric non-Hermitian systems. This is regardless of
whether the bulk spectrum is gapped or gapless, or whether the underlying
Brillouin zone is in standard Bloch or generalized non-Bloch forms.

\section{Summary and Discussion\label{sec:Sum}}

In this work, we revealed and characterized non-Hermiticity induced
critical edge states and topologically nontrivial phase transitions
in 1D non-Hermitian systems with sublattice symmetry. By applying
the Cauchy's argument principle to two characteristic functions of
a non-Hermitian two-band Hamiltonian, we obtain a pair of winding
numbers, whose combination yields an integer-quantized topological
invariant $W$. The applicability of this proposed invariant to the
overall depiction of gapped topological phases, gapless phase boundaries,
topological phase transitions and bulk-edge correspondence is demonstrated
by investigating a broad class of non-Hermitian SSH models. Notably,
we identify a phase transition induced by non-Hermitian effects along
topological phase boundaries, which is accompanied by the quantized
jump of our proposed invariant $W$ and the emergence of critical
edge zero modes. The origin of these nontrivial gapless topology
could not be consistently described by existing approaches relying on
the presence of point or line spectral gaps. One key advantage of
our theoretical approach lies in its generality, in the sense that
it does not concern whether the bulk bands are separated from edge
states of the system by gaps, and whether the underlying
band theory is in Bloch or non-Bloch forms. Therefore, we expect our
theory to be applicable to the topological characterization of phases
and transitions in other more complicated non-Hermitian, two-band
models in one dimension with sublattice symmetry.

One possible way of interpreting the underling physical mechanism of our found gSPT phases may rest on the concept of \emph{symmetry-enriched quantum criticality}~\cite{gSPT14}. We illustrate this point here with an explicit example. The global symmetry that protects the topologically nontrivial critical points of our system is the sublattice symmetry, which reads ${\hat \Gamma}=\sum_n({\hat a}_n^\dagger{\hat a}_n-{\hat b}_n^\dagger{\hat b}_n)$ for NHSSH $\alpha$ chains in the lattice representation. This symmetry enforces an edge zero mode to occupy only one type of sublattice A or B. Therefore, under the OBC, we may identify two sublattice-resolved fermionic parity symmetries at the edges, which are given by ${\hat P}_{1,{\rm A}}=1-2{\hat a}_1^\dagger{\hat a}_1$ and ${\hat P}_{N,{\rm B}}=1-2{\hat b}_N^\dagger{\hat b}_N$, where $N$ is the total number of unit cells. For the NHSSH $(0,1)$ chain, the symmetrized Hamiltonian under the OBC reads ${\hat H}'_{01}={\hat H}'_{01,{\rm Ledge}}+{\hat H}'_{01,{\rm bulk}}+{\hat H}'_{01,{\rm Redge}}$, where ${\hat H}'_{01,{\rm Ledge}}=\varrho^{-1}(J+\gamma){\hat a}_1^{\dagger}{\hat b}_{1}+\varrho(J-\gamma){\hat b}_1^{\dagger}{\hat a}_{1}$, ${\hat H}'_{01,{\rm Redge}}=\varrho^{-1}(J+\gamma){\hat a}_N^{\dagger}{\hat b}_{N}+\varrho(J-\gamma){\hat b}_N^{\dagger}{\hat a}_{N}$, and the ${\hat H}'_{01,{\rm bulk}}$ is given by Eq.~(\ref{eq:H01p}) with the summation taken over $n=1,...,N-1$. At the critical point of the NHSSH $(0,1)$ chain, the ${\hat H}'_{01}$ transforms under the ${\hat P}_{1,{\rm A}}$ and ${\hat P}_{N,{\rm B}}$ as ${\hat P}_{1,{\rm A}}{\hat H}'_{01}{\hat P}_{1,{\rm A}}=-{\hat H}'_{01,{\rm Ledge}}+{\hat H}'_{01,{\rm bulk}}+{\hat H}'_{01,{\rm Redge}}$ and ${\hat P}_{N,{\rm B}}{\hat H}'_{01}{\hat P}_{N,{\rm B}}={\hat H}'_{01,{\rm Ledge}}+{\hat H}'_{01,{\rm bulk}}-{\hat H}'_{01,{\rm Redge}}$. Therefore, the parity symmetries ${\hat P}_{1,{\rm A}}$ and ${\hat P}_{N,{\rm B}}$ are both \emph{broken} at the edges. Their only common eigenstate with the system Hamiltonian is the trivial vacuum state $|\emptyset\rangle$, with ${\hat P}_{1,{\rm A}}|\emptyset\rangle=|\emptyset\rangle$, ${\hat P}_{N,{\rm B}}|\emptyset\rangle=|\emptyset\rangle$, and ${\hat H}'_{01}|\emptyset\rangle=0$. On the other hand, the symmetrized Hamiltonian of NHSSH $(1,2)$ chain under the OBC is given by Eq.~(\ref{eq:H12p}), with the first (second) summation taken over $n=1,...,N-1$ ($n=1,...,N-2$). At its critical point, we have ${\hat P}_{1,{\rm A}}{\hat H}'_{12}{\hat P}_{1,{\rm A}}={\hat P}_{N,{\rm B}}{\hat H}'_{12}{\hat P}_{N,{\rm B}}={\hat H}'_{12}$. Therefore, both the parity symmetries ${\hat P}_{1,{\rm A}}$ and ${\hat P}_{N,{\rm B}}$ are \emph{preserved} at the edges of NHSSH $(1,2)$ chain. Each of these symmetries share a nontrivial eigenstate with ${\hat H}'_{12}$ at one edge, i.e., ${\hat P}_{1,{\rm A}}({\hat a}_1^\dagger|\emptyset\rangle)=-({\hat a}_1^\dagger|\emptyset\rangle)$, ${\hat P}_{N,{\rm B}}({\hat b}_N^\dagger|\emptyset\rangle)=-({\hat b}_N^\dagger|\emptyset\rangle)$, ${\hat H}'_{12}({\hat a}_1^\dagger|\emptyset\rangle)=0({\hat a}_1^\dagger|\emptyset\rangle)$ and ${\hat H}'_{12}({\hat b}_N^\dagger|\emptyset\rangle)=0({\hat b}_N^\dagger|\emptyset\rangle)$. The symmetries ${\hat P}_{1,{\rm A}}$ and ${\hat P}_{N,{\rm B}}$ thus allow two eigenmodes to appear at the two edges of the NHSSH $(1,2)$ chain even when the system is critical. The topological degeneracy of these edge modes at zero energy is further protected by the sublattice symmetry. In comparison to the trivial critical point of the NHSSH $(0,1)$ chain, one may regard the ${\hat P}_{1,{\rm A}}$ and ${\hat P}_{N,{\rm B}}$ as \emph{extra symmetries} that enable the presence of critical edge modes at a topologically nontrivial gapless point. While the explicit form of edge fermionic parity operators could be model-dependent, their existence may serve as a general mechanism to understand the appearance of gSPT phases in other 1D non-Hermitian systems with sublattice symmetry.

As a second remark, we emphasize that it is the sublattice symmetry ${\hat \Gamma}$ that protects the edge modes in the gapped phases and along the gapless phase boundaries in our non-Hermitian system. The ${\hat \Gamma}$ is anti-commute with the system Hamiltonian ${\hat H}$, which means that it enforces energy eigenstates to appear in pairs with opposite signs. To see this, we note that if $|\psi\rangle$ is an eigenstate of ${\hat H}$ with energy $E$, then ${\hat \Gamma}{\hat H}|\psi\rangle={\hat \Gamma}{\hat H}{\hat \Gamma}{\hat \Gamma}|\psi\rangle=-{\hat H}{\hat \Gamma}|\psi\rangle={\hat \Gamma}E|\psi\rangle=E{\hat \Gamma}|\psi\rangle$, so that $H({\hat \Gamma}|\psi\rangle)=-E({\hat \Gamma}|\psi\rangle)$. Therefore, ${\hat \Gamma}|\psi\rangle$ must be an eigenstate of ${\hat H}$ with energy $-E$. If the eigenenergy $E=0$, the states $|\psi\rangle$ and ${\hat \Gamma}|\psi\rangle$ would become degenerate. Therefore, degenerate zero modes of a sublattice-symmetric Hamiltonian must come in pairs. On the other hand, it was known that the topological phase of a 1D, free-fermion lattice model with sublattice symmetry is characterized by an integer-valued topological invariant, which is the $W$ in our theory. In a phase with $W\neq0$ (either gapped or gapless), we will find $N_0=2|W|$ zero-energy eigenmodes at the two edges of the 1D chain in the thermodynamic limit. Each of these edge zero modes has a degenerate partner at zero energy, with their topological degeneracy being protected by the sublattice symmetry. If this phase is gapless, there could also be bulk band touching point at zero energy, whose degeneracy is again enforced by the sublattice symmetry.

Finally, since our theory is applicable to both Hermitian and non-Hermitian cases, the topology it describes does not directly concern whether the critical point is related to a Hermitian-type normal degeneracy or a non-Hermitian exceptional point (EP). In our case studies of NHSSH $(0,1)$ and $(1,2)$ chains, there could be an infinite-order bulk EP within bulk bands at $J=\gamma=\sqrt{2}$ in Figs.~\ref{fig:EP1}(a) and \ref{fig:EP2}(a), where the bulk bands become flat at $|E|=1$. However, in parameter regions at and close to these EPs, the two bulk bands are well gapped at $E=0$, as can be seen from Figs.~\ref{fig:EP1}(a) and \ref{fig:EP2}(a). Therefore, the topology at system edges remains insensitive to these bulk EPs. Overall, even though the EP forms a unique type of non-Hermitian degeneracy, its impact does not show up directly in the topological states we characterized.
	
In future work, it would be interesting to consider the generalization
of our theory to systems with more than two bands \cite{TrimerSSH,Multiband1,Multiband2,Multiband3}, in other symmetry
classes, beyond one spatial dimension, or with impurities \cite{NHimpurity,NHimpurity1,NHimpurity2}. 
Topologically nontrivial criticality and critical edge states that are originated solely from an exceptional-point degeneracy would be intriguing to explore.
The extension of our theory
to driven systems would also be of great importance for revealing
gapless Floquet topology \cite{gSPT34,gSPT35,gSPT36} beyond Hermitian
limits. 
On application side, the topologically protected non-Hermitian criticality may offer alternative routes to the quantum-enhanced sensing in lattice systems \cite{NHSenseRev}.
Finally, the realization of our NHSSH $(0,1,2)$ chain in
quantum simulators like electrical, photonic and acoustic systems \cite{Acoustic1,Acoustic2,EC1,EC2,ECZhao}
may lead to the first experimental observation of critical edge
modes and topologically nontrivial critical points in non-Hermitian
systems, which deserve more thorough explorations.

\begin{acknowledgments}
	This work is supported by the National Natural Science Foundation of China (Grants No.~12275260 and No.~11905211), the Fundamental Research Funds for the Central Universities (Grant No.~202364008), and the Young Talents Project of Ocean University of China.
	
	\emph{Note added}.---After the submission of this manuscript, we noted a preprint arXiv:2509.09587 on non-Hermitian criticality enriched by PT-symmetry, where the bulk theory is deduced with respect to the conventional Brillouin zone of Bloch bands. In our case, the non-Hermitian topological criticality is protected by sublattice symmetry, and the theory is developed with respect to the generalized Brillouin zone of non-Bloch bands, which is also reducible to Bloch band cases in Hermitian limits. 
\end{acknowledgments}
\vspace{0.5cm}

\appendix

\section{Theory: further details}\label{App:Theory}

In this Appendix, we offer additional details and illustrative examples about our theory as sketched in Sec.~\ref{sec:The}.

We first provide some further analyses of the non-Hermitian Hamiltonian in Eq.~(\ref{eq:Hk}).
The off-diagonal elements
$f(k)$ and $g(k)$ in Eq.~(\ref{eq:Hk}) are both $2\pi$-periodic in $k$, and the $H(k)$
is non-Hermitian if there exists a quasimomentum $k$ such that $f(k)\neq g^{*}(k)$.
The sublattice symmetry of $H(k)$ is given by $\Gamma=\sigma_{z}$,
in the sense that $\Gamma H(k)\Gamma=-H(k)$ and $\Gamma^{2}=\sigma_{0}$,
where $\sigma_{0}$ denotes the $2\times2$ identity matrix. The Hamiltonian
$H(k)$ has two energy bands, whose dispersion relations are given
by
\begin{equation}
	E_{\pm}(k)\equiv\pm E(k)=\pm\sqrt{f(k)g(k)}.\label{eq:Ekpm}
\end{equation}
The right eigenvectors of $H(k)$ has the form $\psi(k)\sim[a(k),b(k)]^{\top}$,
where the components $a(k)$ and $b(k)$ satisfy
\begin{equation}
	\frac{b(k)}{a(k)}=\frac{E(k)}{f(k)}=\frac{g(k)}{E(k)}.\label{eq:bova}
\end{equation}
Therefore, all the information about the eigensystem of $H(k)$ are
contained in the ratio $[b(k)/a(k)]^{2}=g(k)/f(k)$ and the product
$E^{2}(k)=f(k)g(k)$. We refer to these two functions as the characteristic
functions of $H(k)$, and express them as
\begin{equation}
	P(k)\equiv{g(k)}/{f(k)},\qquad Q(k)\equiv f(k)g(k).\label{eq:PkQk}
\end{equation}
These functions are also expected to encode the topological properties
and phase transitions of the system described by $H(k)$.

As the functions $f(k)$ and $g(k)$ are both $2\pi$-periodic in $k$, the $P(k)$ and $Q(k)$ in Eq.~(\ref{eq:PkQk}) should also satisfy $P(k+2\pi)=P(k)$
and $Q(k+2\pi)=Q(k)$. We could thus expand the $P(k)$ and $Q(k)$
into Fourier series as $P(k)=\sum_{n\in\mathbb{Z}}p_{n}e^{ink}$ and
$Q(k)=\sum_{n\in\mathbb{Z}}q_{n}e^{ink}$. These series can be viewed
as polynomials of the exponential factor $e^{ik}$. To
treat Bloch and non-Bloch band theories of non-Hermitian systems on
an equal footing, we extend the $e^{ik}$ to the whole complex plane
by setting $e^{ik}\rightarrow z$ with $z\in\mathbb{C}$. The resulting
characteristic polynomials of the complex-extended Hamiltonian $H(z)$
are given by $P(z)=\sum_{n\in\mathbb{Z}}p_{n}z^{n}$ and $Q(z)=\sum_{n\in\mathbb{Z}}q_{n}z^{n}$,
or Eq.~(\ref{eq:PzQz}) in the main text.

In Sec.~\ref{sec:The}, we have mentioned the possible issue that previous theories may encounter at gapless critical points.
The issue can be made clear via a counterexample. Let us consider
a complex-continued non-Hermitian Hamiltonian $H(z)$ with off-diagonal
elements $f(z)=J_{0}+J_{1}z^{-1}$ and $g(z)=J_{0}+J_{1}z$. When
$J_{0}$ and $J_{1}$ take real values, we go back to the standard
SSH model after setting $z\rightarrow e^{ik}$. Following the $P(z)$-based
approach as discussed in Sec.~\ref{sec:The}, we have $m=1$, $n=0$ and thus $2\mu=2\max\{1,0\}=2$.
By definition, we find $P(z)=g(z)/f(z)=z(J_{0}+J_{1}z)/(J_{0}z+J_{1})$.
It can then be identified that if $|J_{1}|<|J_{0}|$, the first $2$
zeros/poles of $P(z)$ contains a zero at $z=0$ and a pole at $z=-J_{1}/J_{0}$,
yielding $N_{z}^{\in2}=1$, $N_{p}^{\in2}=1$, $W_{0}=N_{z}^{\in2\mu}-N_{p}^{\in2\mu}=0$
and $N_{0}=|W_{0}|=0$. Meanwhile, if $|J_{1}|>|J_{0}|$, the first
$2$ zeros/poles of $P(z)$ contains two zeros at $z=0$ and $z=-J_{0}/J_{1}$,
yielding $N_{z}^{\in2}=2$, $N_{p}^{\in2}=0$, $W_{0}=N_{z}^{\in2\mu}-N_{p}^{\in2\mu}=2$
and $N_{0}=|W_{0}|=2$. These predictions are all correct, as the
two bulk bands of $H(z)$ are gapped at zero energy {[}$Q(z)\neq0${]}
for any finite $z$ whenever $|J_{1}|\neq|J_{0}|$. However, at the
phase transition point $|J_{1}|=|J_{0}|$, we have $|-J_{1}/J_{0}|=|-J_{0}/J_{1}|=1$,
which means that we could not order the magnitudes of the zero and
pole at $z=-J_{0}/J_{1}$ and $z=-J_{1}/J_{0}$. The determination
of which are the first two zeros and/or poles of $P(z)$ then becomes
ambiguous. If we include or exclude both their contributions to $W_{0}$,
we will get $W_{0}=1$ and thus $N_{0}=1$. This is impossible,
as the zero modes must come in pairs (i.e., $N_{0}\in2\mathbb{N}$)
due to the sublattice symmetry. If we only retain the zero at $z=-J_{0}/J_{1}$
as the second zero/pole, we will get $W_{0}=2$ and $N_{0}=2$, which
is incorrect even for the Hermitian SSH model \cite{gSPT03}. If we
instead retain the pole at $z=-J_{1}/J_{0}$ as the second zero/pole,
we will find $W_{0}=0$ and $N_{0}=0$. Even though these are the
correct values of topological invariants and edge states at
the critical point of SSH model \cite{gSPT03}, they could
not be conclusively and unambiguously predicted by the $P(z)$-based
approach. Therefore, the algebraic scheme in Refs.~\cite{NHSE04,NHSE07}
tends out to be inapplicable to the characterization of nontrivial
topology and bulk-edge correspondence at the phase transition (bulk-gap
closing) point of non-Hermitian systems. Similarly, as the algebraic
and GBZ schemes are proved to be equivalent for sublattice-symmetric
models \cite{NHSE07}, the standard GBZ approach is also expected
to be unworkable for describing gSPT phases in non-Hermitian systems.

These observations motivate us to generalize existing theories in order to
characterize the nontrivial topology at non-Hermitian critical
points. The above analysis suggests that for the SSH model, there could be competing zeros at $z=-J_{0}/J_{1}$
and poles at $z=-J_{1}/J_{0}$ of equal magnitudes in $P(z)$ at the critical point. Their contributions
are encoded in $Q(z)$ as
its zeros at the gapless point. Therefore, we introduce
another winding number by applying the Cauchy's argument principle
to $Q(z)$. The winding numbers of $P(z)$ and $Q(z)$
are finally given by Eq.~(\ref{eq:W12}) of the main text.

We may now check the applicability of our theory to the
SSH model. Besides the $P(z)=z(J_{0}+J_{1}z)/(J_{0}z+J_{1})$,
we also have $Q(z)=(J_{0}z+J_{1})(J_{0}+J_{1}z)/z$. The GBZ here
is just the standard Brillouin zone, which forms a circle of radius
$1$ on the complex plane regardless of whether the values
of $J_{0}$ and $J_{1}$ are real or complex. Direct calculations
following Eqs.~(\ref{eq:W12})--(\ref{eq:N0}) in the main text predict $(W_{1},W_{2})=(0,0)$
when $|J_{1}|<|J_{0}|$, $(W_{1},W_{2})=(2,0)$ when $|J_{1}|>|J_{0}|$,
and $(W_{1},W_{2})=(1,-1)$ when $|J_{1}|=|J_{0}|$, yielding $W=0,1,0$
and $N_{0}=0,2,0$ in these three parameter regions. Besides reproducing
known results about gapped phases~\cite{TPBook}, we are now able to confirm
the topological triviality of the gapless phase boundary $|J_{1}|=|J_{0}|$
of the SSH model~\cite{gSPT04} without ambiguity.

\section{Single $\alpha$ chain: gapped and gapless phases}\label{sec:SSHa}

In this Appendix, we investigate the simplest type of NHSSH $\alpha$ chain, whose Hamiltonian is given by ${\hat H}={\hat H}_\alpha$ for a fixed $\alpha$. This NHSSH single $\alpha$ chain has the complex-extended Hamiltonian $H(z)=H_{\alpha}(z)$. It contains only a single length scale in its hopping amplitudes $J_{\alpha}^{{\rm L,R}}$. Throughout this Appendix, we assume $\alpha\geq0$ without losing generality.

Following the theoretical development in Sec.~\ref{sec:The}, we find
$f(z)=J_{\alpha}^{{\rm R}}z^{-\alpha}$ and $g(z)=J_{\alpha}^{{\rm L}}z^{\alpha}$
for the single $\alpha$ chain. The characteristic functions
are thus given by
\begin{equation}
	P(z)=\frac{J_{\alpha}^{{\rm L}}}{J_{\alpha}^{{\rm R}}}z^{2\alpha},\qquad Q(z)=J_{\alpha}^{{\rm L}}J_{\alpha}^{{\rm R}}.\label{eq:PQaz}
\end{equation}
Since the $Q(z)=[E(z)]^{2}$ is independent of $z$, the bulk spectrum
of the system is formed by two flat bands at $\pm E(z)=\pm\sqrt{J_{\alpha}^{{\rm L}}J_{\alpha}^{{\rm R}}}$,
and the GBZ is just the standard BZ with $k\in[-\pi,\pi]$. On the
complex $z$-plane, the $Q(z)$ has neither zeros nor poles, while
the $P(z)$ has a zero of order $2\alpha$ at $z=0$, which is inside
the BZ unit circle $|z|=1$. According to the definition of winding
numbers for $P(z)$ and $Q(z)$ in Eq.~(\ref{eq:W12}), we find $W_{1}=2\alpha$
and $W_{2}=0$. The topological invariant $W$ of the system and the
number of its degenerate zero modes $N_{0}$ under the OBC are thus given by Eqs.~(\ref{eq:W}) and (\ref{eq:N0})
as
\begin{equation}
	W=\frac{W_{1}+W_{2}}{2}=\alpha,\quad N_{0}=2|W|=2\alpha.\label{eq:WBBCa}
\end{equation}
We find that the topological phase of the system is independent of
its hopping amplitudes, so long as the $J_{\alpha}^{{\rm L}}$ and
$J_{\alpha}^{{\rm R}}$ are not simultaneously zero. Meanwhile, the
bulk spectrum of the system is gapped (gapless) when $J_{\alpha}^{{\rm L}}\neq0$
and $J_{\alpha}^{{\rm R}}\neq0$ ($J_{\alpha}^{{\rm L}}=0$ or $J_{\alpha}^{{\rm R}}=0$).
The NHSSH single $\alpha$ chain could thus admit both gapped and
gapless topologically nontrivial phases whenever $\alpha\neq0$.

We now verify our results about the NHSSH single $\alpha$ chain with
two groups of numerical examples. In the first group, we consider
the case of asymmetric hopping with $(J_{\alpha}^{{\rm L}},J_{\alpha}^{{\rm R}})=(0.5,2)$
for $\alpha=0,1,2$. The bulk spectrum is expected to show two flat
bands at $\pm\sqrt{J_{\alpha}^{{\rm L}}J_{\alpha}^{{\rm R}}}=\pm1$,
separated by a constant gap $\Delta E=2$. 
In the second group, we
consider the case of unidirectional hopping with $(J_{\alpha}^{{\rm L}},J_{\alpha}^{{\rm R}})=(0,1)$
for $\alpha=0,1,2$. The bulk spectrum is expected to show two degenerate
flat bands at zero energy. In these cases, the theoretically
predicted winding numbers and numbers of zero-energy edge modes are
$W=(0,1,2)$ and $N_{0}=(0,2,4)$ according to Eq.~(\ref{eq:WBBCa}),
regardless of the chosen parameter values $(J_{\alpha}^{{\rm L}},J_{\alpha}^{{\rm R}})$.

\begin{figure}
	\begin{centering}
		\includegraphics[scale=0.5]{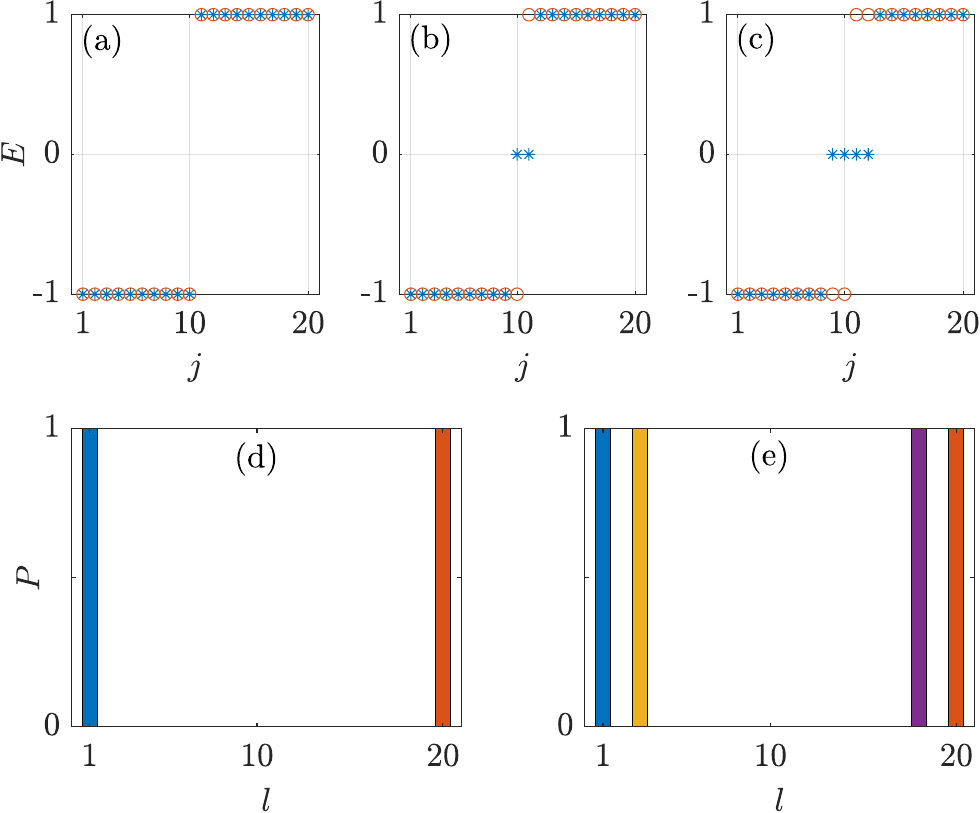}
		\par\end{centering}
	\caption{Energy spectra $E$ and edge states of NHSSH single $\alpha$ chains with
		bulk gaps. The system has $10$ unit cells and $20$ lattice sites.
		$j$ is the state index. $l$ is the lattice index. (a) Spectrum of $\hat{H}_{0}$ with $(J_{0}^{{\rm L}},J_{0}^{{\rm R}})=(0.5,2)$
		under the PBC (in circles) and OBC (in stars). (b) Spectrum
		of $\hat{H}_{1}$ with $(J_{1}^{{\rm L}},J_{1}^{{\rm R}})=(0.5,2)$
		under the PBC (in circles) and OBC (in stars). (c) Spectrum
		of $\hat{H}_{2}$ with $(J_{2}^{{\rm L}},J_{2}^{{\rm R}})=(0.5,2)$
		under the PBC (in circles) and OBC (in stars). (d) Probability distributions
		$P$ of the zero-energy edge states of $\hat{H}_{1}$ in (b).
		(e) Probability distributions $P$ of the zero-energy
		edge states of $\hat{H}_{2}$ in (c).
		In (d) and (e), different edge states are distinguished
		by the colors of probability bars. \label{fig:SAC1}}
\end{figure}

In Fig.~\ref{fig:SAC1}, we present the energy spectra and edge states
of the single $\alpha$ chain for $\alpha=0,1,2$ and $(J_{\alpha}^{{\rm L}},J_{\alpha}^{{\rm R}})=(0.5,2)$.
The spectrum of $\hat{H}_{0}$ in Fig.~\ref{fig:SAC1}(a) has no signatures
of zero-energy edge modes ($N_{0}=0$). The spectrum of $\hat{H}_{1}$
in Fig.~\ref{fig:SAC1}(b) possesses two zero-energy edge modes ($N_{0}=2$),
whose probability distributions are shown in Fig.~\ref{fig:SAC1}(d).
The spectrum of $\hat{H}_{2}$ in Fig.~\ref{fig:SAC1}(c) has four
zero-energy edge modes ($N_{0}=4$), whose probability distributions
are shown in Fig.~\ref{fig:SAC1}(e). These numerical results are
all consistent with our theoretical predictions.

\begin{figure}
	\begin{centering}
		\includegraphics[scale=0.5]{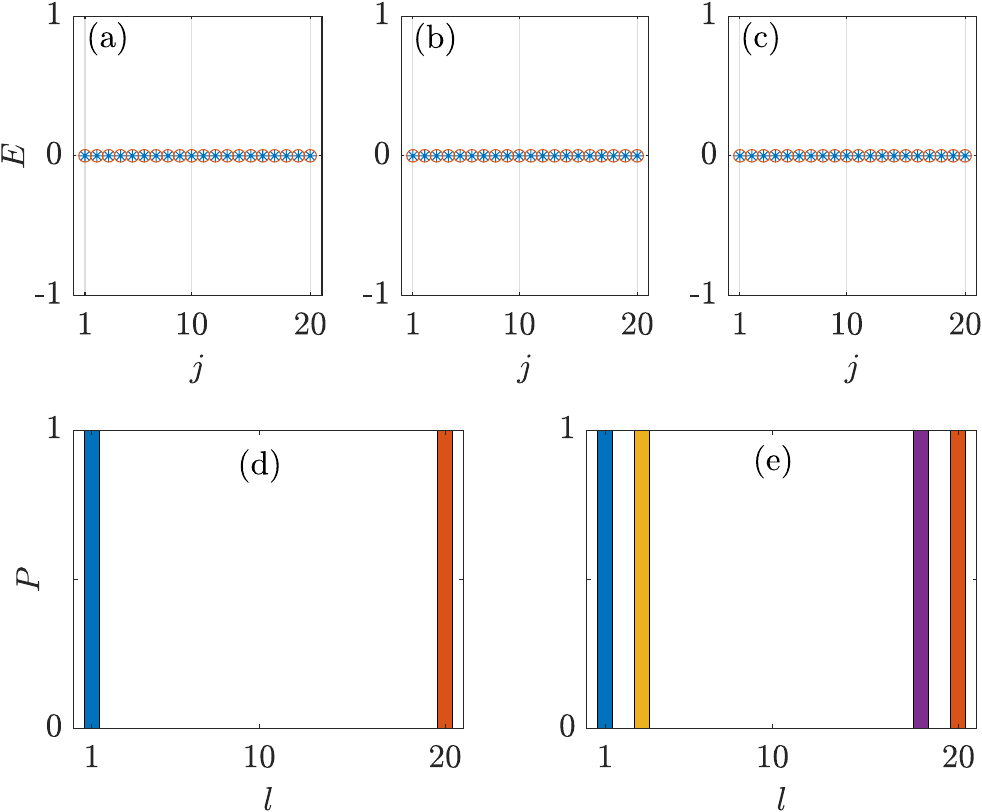}
		\par\end{centering}
	\caption{Energy spectra $E$ and edge states of NHSSH single $\alpha$ chains without
		bulk gaps. The system has $10$ unit cells and $20$ lattice sites.
		$j$ is the state index. $l$ is the lattice index. (a) Spectrum
		of $\hat{H}_{0}$ with $(J_{0}^{{\rm L}},J_{0}^{{\rm R}})=(0,1)$
		under the PBC (in circles) and OBC (in stars). (b) Spectrum
		of $\hat{H}_{1}$ with $(J_{1}^{{\rm L}},J_{1}^{{\rm R}})=(0,1)$
		under the PBC (in circles) and OBC (in stars). (c) Spectrum
		of $\hat{H}_{2}$ with $(J_{2}^{{\rm L}},J_{2}^{{\rm R}})=(0,1)$
		under the PBC (in circles) and OBC (in stars). (d) Probability distributions
		$P$ of the zero-energy edge states of $\hat{H}_{1}$ in (b).
		(e) Probability distributions $P$ of the zero-energy
		edge states of $\hat{H}_{2}$ in (c).
		In (d) and (e), different edge states are distinguished
		by the colors of probability bars. \label{fig:SAC2}}
\end{figure}

In Fig.~\ref{fig:SAC2}, we present the energy spectra and edge states
of the single $\alpha$ chain for $\alpha=0,1,2$ and $(J_{\alpha}^{{\rm L}},J_{\alpha}^{{\rm R}})=(0,1)$.
The spectrum of $\hat{H}_{0}$ in Fig.~\ref{fig:SAC2}(a) has no edge
zero modes ($N_{0}=0$). The spectrum of $\hat{H}_{1}$ in Fig.~\ref{fig:SAC2}(b)
holds two zero-energy edge modes ($N_{0}=2$), whose distributions
are shown in Fig.~\ref{fig:SAC2}(d). The spectrum of $\hat{H}_{2}$
in Fig.~\ref{fig:SAC2}(c) contains four zero-energy edge modes ($N_{0}=4$),
whose distributions are shown in Fig.~\ref{fig:SAC2}(e). These results
again verify our theoretical predictions. Notably, even though the
bulk spectra of $\hat{H}_{0,1,2}$ have no gaps at $E=0$ in Figs.~\ref{fig:SAC2}(a)--\ref{fig:SAC2}(c), our theory could still correctly
predict the numbers of edge zero modes that can coexist with gapless
bulks in these cases.
Note that for our models at half-filling, gaplessness means that the two bulk bands are touched at least at one point in the energy spectrum. This is regardless of whether the two bands are presented in momentum or lattice representation. Since the spectra in Figs.~\ref{fig:SAC2}(a--c) are obtained under both the periodic and open boundary conditions in lattice representation, we do not have a conserved quasimomentum $k$ in both cases. The spectra are thus shown with respect to the index of energy eigenstate there.

The Figs.~\ref{fig:SAC2}(a--c) are obtained for a special choice of system parameters, under which the two bulk bands of the system are completely degenerate at zero energy. One may also see this from the Bloch Hamiltonian of the bulk under periodic boundary condition, which in this case is given by $H_\alpha(k)=J_\alpha^{\rm R}(\sigma_x+i\sigma_y)e^{-ik\alpha}/2$. It has the dispersion relation $E_\pm(k)=0$ for any $\alpha$. Therefore, the two bulk bands are both flat and they are degenerate at every $k\in[-\pi,\pi)$. The bulk spectrum of the system is further gapless in this case for any $J_\alpha^{\rm R}$, thus realizing a gapless phase in the parameter space $J_\alpha^{\rm R}\in{\mathbb C}$.

The presence or absence of edge zero modes in the spectrum of $\hat{H}_{\alpha}$
for a given $\alpha$ can be understood from the geometric
connectivity of the underlying lattice [see Fig.~\ref{fig:Sketch}(c)].
For $\hat{H}_{0}$, the sublattices A and B in each unit cell are
coupled due to the non-vanishing $J_{0}^{{\rm L}}$ and/or $J_{0}^{{\rm R}}$,
making it impossible to have isolated sites at boundaries to host
edge zero modes. For $\hat{H}_{1}$, the sublattice B in unit cell
$n$ is coupled to the sublattice A in unit cell $n+1$ for $n=1,...,N-1$.
The sublattices A in unit cell $1$ and B in unit cell $N$ are thus
isolated from other sites, making them available to host two
edge zero modes $\hat{a}_{1}^{\dagger}|\emptyset\rangle$ and $\hat{b}_{N}^{\dagger}|\emptyset\rangle$,
where $|\emptyset\rangle$ denotes the vacuum state. For $\hat{H}_{2}$,
the sublattice B in unit cell $n$ is coupled to the sublattice A
in unit cell $n+2$ for $n=1,...,N-2$. The sublattices A in unit
cells $1,2$ and B in unit cells $N-1,N$ are thus isolated from other
sites, making them available to host four edge zero modes $\hat{a}_{1}^{\dagger}|\emptyset\rangle$,
$\hat{a}_{2}^{\dagger}|\emptyset\rangle$, $\hat{b}_{N-1}^{\dagger}|\emptyset\rangle$
and $\hat{b}_{N}^{\dagger}|\emptyset\rangle$. It is also clear that
the topological origin of zero-energy edge modes in a single $\alpha$
chain does not depend on whether the system is Hermitian or non-Hermitian.
Note that for the $\alpha=1$ and $\alpha=2$ chains, their edge zero modes are robust to any perturbations (such as disorder) added to the hopping amplitudes $(J_\alpha^{\rm L},J_\alpha^{\rm R})$ over different unit cells, as these perturbations could not couple the isolated edge zero modes of these single $\alpha$ chains to bulk states. The topological degeneracy of these edge zero modes is protected the sublattice symmetry of the bulk.

\section{Calculation of the edge states\label{sec:App1}}

In this Appendix, we compute the edge modes analytically and
determine their existent conditions for NHSSH $(\alpha,\alpha')$
chains with $(\alpha,\alpha')=(0,1)$ and $(1,2)$.

In lattice representation, the Hamiltonian of NHSSH $(0,1)$
chain in Sec.~\ref{subsec:SSHaa} takes the form
\begin{alignat}{1}
	\hat{H}_{01}= & \sum_{n\in\mathbb{Z}}\left[(J+\gamma)\hat{a}_{n}^{\dagger}\hat{b}_{n}+(J-\gamma)\hat{b}_{n}^{\dagger}\hat{a}_{n}\right]\nonumber \\
	+ & \sum_{n\in\mathbb{Z}}\left(\hat{b}_{n}^{\dagger}\hat{a}_{n+1}+\hat{a}_{n+1}^{\dagger}\hat{b}_{n}\right).\label{eq:H01}
\end{alignat}
Following Ref.~\cite{NHSE02}, we apply a similarity transformation
$\hat{S}$ to $\hat{H}_{01}$ under the OBC, where
\begin{equation}
	\hat{S}=\sum_{n\in\mathbb{Z}}\left(\varrho^{n-1}\hat{a}_{n}^{\dagger}\hat{a}_{n}+\varrho^{n}\hat{b}_{n}^{\dagger}\hat{b}_{n}\right).\label{eq:S}
\end{equation}
$\varrho$ is a control parameter. The transformed Hamiltonian takes
the form $\hat{H}'_{01}=\hat{S}\hat{H}_{01}\hat{S}^{-1}$, where 
\begin{alignat}{1}
	\hat{H}'_{01}= & \sum_{n\in\mathbb{Z}}\left[\varrho^{-1}(J+\gamma)\hat{a}_{n}^{\dagger}\hat{b}_{n}+\varrho(J-\gamma)\hat{b}_{n}^{\dagger}\hat{a}_{n}\right]\nonumber \\
	+ & \sum_{n\in\mathbb{Z}}\left(\hat{b}_{n}^{\dagger}\hat{a}_{n+1}+\hat{a}_{n+1}^{\dagger}\hat{b}_{n}\right).\label{eq:H01p}
\end{alignat}
We choose $\varrho$ to ensure the left and right intracell hoppings
to have equal magnitudes, which means that $|\rho|=\sqrt{|J+\gamma|/|J-\gamma|}$.
Let us consider a chain with unit cell indices $n=1,2,...,N$ and
take the OBC at both ends. A zero-energy eigenmode of $\hat{H}'_{01}$
satisfies $\hat{H}'_{01}|\psi_{0}\rangle=0$. Expanding $|\psi_{0}\rangle$
in the lattice representation as $|\psi_{0}\rangle=\sum_{m}(A_{m}\hat{a}_{m}^{\dagger}+B_{m}\hat{b}_{m}^{\dagger})|\emptyset\rangle$
and applying $\hat{H}'_{01}$
to it, we get the iteration equations of wave amplitudes
$(A_{n},B_{n})$, i.e.,
\begin{equation}
	\varrho^{-1}(J+\gamma)B_{1}=0,\qquad\varrho(J-\gamma)A_{N}=0,
\end{equation}
\begin{equation}
	\varrho(J-\gamma)A_{n}+A_{n+1}=0,\qquad n=1,...,N-1,
\end{equation}
\begin{equation}
	\varrho^{-1}(J+\gamma)B_{n+1}+B_{n}=0,\qquad n=1,...,N-1.
\end{equation}
In the limit $N\rightarrow\infty$, this set of difference equation
has two zero-energy solutions. Their wave functions (up to normalization
factors) are given by
\begin{alignat}{1}
	|\psi_{0}^{{\rm L}}\rangle= & \sum_{n=1}^{N}[-\varrho(J-\gamma)]^{n-1}\hat{a}_{n}^{\dagger}|\emptyset\rangle,\nonumber \\
	|\psi_{0}^{{\rm R}}\rangle= & \sum_{n=1}^{N}[-(J+\gamma)/\varrho]^{N-n}\hat{b}_{n}^{\dagger}|\emptyset\rangle.\label{eq:0aa}
\end{alignat}
It is clear that the states $|\psi_{0}^{{\rm L}}\rangle$ and $|\psi_{0}^{{\rm R}}\rangle$
represent left-localized and right-localized edge zero modes if
and only if $|\varrho(J-\gamma)|<1$ and $|(J+\gamma)/\varrho|<1$,
respectively. Both these conditions yield $|J^{2}-\gamma^{2}|<1$,
which is exactly the condition for us to have topologically nontrivial
gapped phases with the winding number $W=1$ and the number of zero-energy
edge modes $N_{0}=2$, as reported in Eqs.~(\ref{eq:Waa1}) and (\ref{eq:BBCaa1}).
The bulk-edge correspondence of our NHSSH $(0,1)$ chain is thus proved.
Importantly, we find no zero-energy edge modes when $J^{2}-\gamma^{2}=\pm1$,
i.e., at the critical points of the system under OBC. This observation
confirms that the gapless phase boundary of the NHSSH $(0,1)$ chain
is indeed topologically trivial from the perspective of critical edge
states.

We next consider the NHSSH $(1,2)$ chain in Sec.~\ref{subsec:SSHaa},
whose Hamiltonian takes the form
\begin{alignat}{1}
	\hat{H}_{12}= & \sum_{n\in\mathbb{Z}}\left[(J-\gamma)\hat{b}_{n}^{\dagger}\hat{a}_{n+1}+(J+\gamma)\hat{a}_{n+1}^{\dagger}\hat{b}_{n}\right]\nonumber \\
	+ & \sum_{n\in\mathbb{Z}}\left(\hat{b}_{n}^{\dagger}\hat{a}_{n+2}+\hat{a}_{n+2}^{\dagger}\hat{b}_{n}\right).\label{eq:H12}
\end{alignat}
Under a similarity transformation
\begin{equation}
	\hat{R}=\sum_{n\in\mathbb{Z}}\left(\varrho^{n-1}\hat{a}_{n}^{\dagger}\hat{a}_{n}+\varrho^{n+1}\hat{b}_{n}^{\dagger}\hat{b}_{n}\right),\label{eq:R}
\end{equation}
the $\hat{H}_{12}$ becomes $\hat{H}'_{12}=\hat{R}\hat{H}_{12}\hat{R}^{-1}$,
or
\begin{alignat}{1}
	\hat{H}'_{12}= & \sum_{n\in\mathbb{Z}}\left[\varrho(J-\gamma)\hat{b}_{n}^{\dagger}\hat{a}_{n+1}+\varrho^{-1}(J+\gamma)\hat{a}_{n+1}^{\dagger}\hat{b}_{n}\right]\nonumber \\
	+ & \sum_{n\in\mathbb{Z}}\left(\hat{b}_{n}^{\dagger}\hat{a}_{n+2}+\hat{a}_{n+2}^{\dagger}\hat{b}_{n}\right).\label{eq:H12p}
\end{alignat}
We choose $\varrho$ to let the left and right intracell hoppings
to have equal magnitudes, which again leads to $|\rho|=\sqrt{|J+\gamma|/|J-\gamma|}$.
For a chain with unit cell indices $n=1,2,...,N$ and under the OBC
at both ends, a zero-energy solution of $\hat{H}'_{12}$ satisfies
$\hat{H}'_{12}|\varphi_{0}\rangle=0$. It is not hard to notice that
one set of solutions of this equation is given by the eigenmodes
\begin{equation}
	|\varphi_{0}^{{\rm L,1}}\rangle=\hat{a}_{1}^{\dagger}|\emptyset\rangle,\qquad|\varphi_{0}^{{\rm R,1}}\rangle=\hat{b}_{N}^{\dagger}|\emptyset\rangle,\label{eq:0aa2}
\end{equation}
which form a pair of degenerate edge modes at the left and right
ends of the lattice. To find other possible solutions, we again expand
the zero mode in the lattice space as $|\varphi_{0}\rangle=\sum_{m}(C_{m}\hat{a}_{m}^{\dagger}+D_{m}\hat{b}_{m}^{\dagger})|\emptyset\rangle$
and apply the Hamiltonian $\hat{H}'_{12}$ to it. The resulting zero-energy
solutions in the limit $N\rightarrow\infty$ are given by
\begin{alignat}{1}
	|\varphi_{0}^{{\rm L,2}}\rangle= & \sum_{n=2}^{N}[-\varrho(J-\gamma)]^{n-2}\hat{a}_{n}^{\dagger}|\emptyset\rangle,\nonumber \\
	|\varphi_{0}^{{\rm R,2}}\rangle= & \sum_{n=1}^{N-1}[-(J+\gamma)/\varrho]^{N-n-1}\hat{b}_{n}^{\dagger}|\emptyset\rangle.\label{eq:0aa3}
\end{alignat}
The states $|\varphi_{0}^{{\rm L,2}}\rangle$ and $|\varphi_{0}^{{\rm R,2}}\rangle$
represent left-localized and right-localized eigenmodes if and only
if $|\varrho(J-\gamma)|<1$ and $|\varrho^{-1}(J+\gamma)|<1$, respectively,
yielding the condition $|J^{2}-\gamma^{2}|<1$ for their existence.
This is the same condition for us to have topologically nontrivial
gapped phases with the winding number $W=2$ and the number of
edge zero modes $N_{0}=4$, as reported in Eqs.~(\ref{eq:Waa2}) and (\ref{eq:BBCaa2}).
Since the edge modes $|\varphi_{0}^{{\rm L,1}}\rangle$ and $|\varphi_{0}^{{\rm R,1}}\rangle$
are persistent under this condition, the bulk-edge correspondence
of the $W=2$ phase is confirmed. When $|J^{2}-\gamma^{2}|>1$, the
edge zero modes $|\varphi_{0}^{{\rm L,2}}\rangle$ and $|\varphi_{0}^{{\rm R,2}}\rangle$
disappear, while the modes $|\varphi_{0}^{{\rm L,1}}\rangle$ and
$|\varphi_{0}^{{\rm R,1}}\rangle$ are retained. This conforms to
our expectation for the other gapped topological phase with the winding
number $W=1$ and the number of edge zero modes $N_{0}=2$, as shown
in Eqs.~(\ref{eq:Waa2}) and (\ref{eq:BBCaa2}). Finally, we notice
that the edge modes $|\varphi_{0}^{{\rm L,1}}\rangle$ and $|\varphi_{0}^{{\rm R,1}}\rangle$
survive at the critical points $|J^{2}-\gamma^{2}|=1$. Therefore,
the phase boundary of the NHSSH $(1,2)$ chain is topologically
nontrivial, which is characterized by a quantized winding
number $W=1$ and a pair of edge modes degenerating with a gapless
bulk at zero energy.

\end{document}